\newcommand\apjcls{1}
\newcommand\aastexcls{2}
\newcommand\othercls{3}

\newcommand\papercls{\aastexcls}
\documentclass[tighten, times,  preprint2]{aastex6} 

\if\papercls \apjcls
\usepackage{apjfonts}
\else\if\papercls \othercls
\usepackage{epsfig}
\fi\fi
\usepackage{ifthen}
\usepackage{natbib}
\usepackage{amssymb, amsmath}
\usepackage{appendix}
\usepackage{etoolbox}

\if\papercls \apjcls
\newcommand\aas{\ref@jnl{AAS Meeting Abstracts}}
\newcommand\dps{\ref@jnl{AAS/DPS Meeting Abstracts}}
\newcommand\maps{\ref@jnl{MAPS}}
\else\if\papercls \othercls
\usepackage{astjnlabbrev-jh}
\fi\fi

\if\papercls \apjcls
\bibpunct[, ]{(}{)}{,}{a}{}{,}
\else
\bibpunct[, ]{(}{)}{,}{a}{}{,}
\fi

\if\papercls \aastexcls
\hypersetup{citecolor=blue, 
            linkcolor=blue, 
            menucolor=blue, 
            urlcolor=blue}  
\else
\usepackage[
bookmarks=true,           
bookmarksnumbered=true,   
colorlinks=true,          
citecolor=blue,           
linkcolor=blue,           
menucolor=blue,           
urlcolor=blue,            
linkbordercolor={0 0 1},  
pdfborder={0 0 1},
frenchlinks=true]{hyperref}
\fi

\if\papercls \othercls
\newcommand{\eprint}[1]{\href{http://arxiv.org/abs/#1}{#1}}
\else
\renewcommand{\eprint}[1]{\href{http://arxiv.org/abs/#1}{#1}}
\fi
\newcommand{\ISBN}[1]{\href{http://cosmologist.info/ISBN/#1}{ISBN: #1}}
\providecommand{\adsurl}[1]{\href{#1}{ADS}}

\makeatletter
\patchcmd{\NAT@citex}
  {\@citea\NAT@hyper@{%
     \NAT@nmfmt{\NAT@nm}%
     \hyper@natlinkbreak{\NAT@aysep\NAT@spacechar}{\@citeb\@extra@b@citeb}%
     \NAT@date}}
  {\@citea\NAT@nmfmt{\NAT@nm}%
   \NAT@aysep\NAT@spacechar\NAT@hyper@{\NAT@date}}{}{}

\patchcmd{\NAT@citex}
  {\@citea\NAT@hyper@{%
     \NAT@nmfmt{\NAT@nm}%
     \hyper@natlinkbreak{\NAT@spacechar\NAT@@open\if*#1*\else#1\NAT@spacechar\fi}%
       {\@citeb\@extra@b@citeb}%
     \NAT@date}}
  {\@citea\NAT@nmfmt{\NAT@nm}%
   \NAT@spacechar\NAT@@open\if*#1*\else#1\NAT@spacechar\fi\NAT@hyper@{\NAT@date}}
  {}{}
\makeatother

\makeatletter
\DeclareRobustCommand{\lowcase}[1]{\@lowcase#1\@nil}
\def\@lowcase#1\@nil{\if\relax#1\relax\else\MakeLowercase{#1}\fi}
\makeatother

\DeclareSymbolFont{UPM}{U}{eur}{m}{n}
\DeclareMathSymbol{\umu}{0}{UPM}{"16}
\let\oldumu=\umu
\renewcommand\umu{\ifmmode\oldumu\else\math{\oldumu}\fi}
\newcommand\micro{\umu}
\if\papercls \othercls
\newcommand\micron{\micro m}
\else
\renewcommand\micron{\micro m}
\fi
\newcommand\microns{\micron}

\let\oldsim=\sim
\renewcommand\sim{\ifmmode\oldsim\else\math{\oldsim}\fi}
\let\oldpm=\pm
\renewcommand\pm{\ifmmode\oldpm\else\math{\oldpm}\fi}
\newcommand\by{\ifmmode\times\else\math{\times}\fi}

\newcommand\tablebox[1]{\begin{tabular}[t]{@{}l@{}}#1\end{tabular}}
\newbox{\wdbox}
\renewcommand\c{\setbox\wdbox=\hbox{,}\hspace{\wd\wdbox}}
\renewcommand\i{\setbox\wdbox=\hbox{i}\hspace{\wd\wdbox}}

\newcount\timect
\newcount\hourct
\newcount\minct
\newcommand\now{\timect=\time \divide\timect by 60
         \hourct=\timect \multiply\hourct by 60
         \minct=\time \advance\minct by -\hourct
         \number\timect:\ifnum \minct < 10 0\fi\number\minct}

\catcode`@=11

\newcommand\comment[1]{}

\newcommand\commenton{\catcode`\%=14}

\renewcommand\math[1]{$#1$}
\newcommand\mathshifton{\catcode`\$=3}

\let\atab=&
\newcommand\atabon{\catcode`\&=4}

\let\oldmsp=\sp
\let\oldmsb=\sb
\def\sp#1{\ifmmode
           \oldmsp{#1}%
         \else\strut\raise.85ex\hbox{\scriptsize #1}\fi}
\def\sb#1{\ifmmode
           \oldmsb{#1}%
         \else\strut\raise-.54ex\hbox{\scriptsize #1}\fi}
\newbox\@sp
\newbox\@sb
\def\sbp#1#2{\ifmmode%
           \oldmsb{#1}\oldmsp{#2}%
         \else
           \setbox\@sb=\hbox{\sb{#1}}%
           \setbox\@sp=\hbox{\sp{#2}}%
           \rlap{\copy\@sb}\copy\@sp
           \ifdim \wd\@sb >\wd\@sp
             \hskip -\wd\@sp \hskip \wd\@sb
           \fi
        \fi}
\def\msp#1{\ifmmode
           \oldmsp{#1}
         \else \math{\oldmsp{#1}}\fi}
\def\msb#1{\ifmmode
           \oldmsb{#1}
         \else \math{\oldmsb{#1}}\fi}

\def\supon{\catcode`\^=7}

\def\subon{\catcode`\_=8}

\def\supsubon{\supon \subon}

\newcommand\actcharon{\catcode`\~=13}

\newcommand\paramon{\catcode`\#=6}

\comment{And now to turn us totally on and off...}

\newcommand\reservedcharson{ \commenton  \mathshifton  \atabon  \supsubon 
                             \actcharon  \paramon}

\catcode`@=12
\reservedcharson

\if\papercls \apjcls

\else

\fi

\if\papercls \othercls
\else
  \newcommand\inpress{n}
  \if\inpress y
    \received{\today}
    \revised{}
    \accepted{}
    \if\papercls \apjcls
    \slugcomment{}
    \fi
  \else
  \slugcomment{\tablebox{In preparation for {\em ApJ}. DRAFT of {\today}.}}
  \fi
\fi

\newcommand\chisq{\ifmmode{\chi\sp{2}}\else\math{\chi\sp{2}}\fi}
\newcommand\redchisq{\ifmmode{ \chi\sp{2}\sb{\rm red}}
                    \else\math{\chi\sp{2}\sb{\rm red}}\fi}

\newcommand\mcc{MC\sp{3}}

\newcommand\vs{\emph{vs.}}

\newcommand{\ceq}[1]{\eqref{#1}}
\newcommand{\yvec}{\vec{y}}

\newcommand{\fvec}{\vec{f}}

\newcommand{\evec}{\vec{\epsilon}}

\newcommand{\params}{\theta}

\newcommand{\like}{\mathcal{L}}
\newcommand{\eps}{\epsilon}

\newcommand\carter{CW09}
\defcitealias{CarterWinn2009apjWavelets}{\carter}

\shorttitle{On Correlated-Noise Analyses}
\shortauthors{Cubillos {\em et al.}}

\begin{document}

\title{On Correlated-Noise Analyses Applied To Exoplanet Light Curves}

\author{Patricio~Cubillos\altaffilmark{1,2},
Joseph~Harrington\altaffilmark{1},
Thomas~J.~Loredo\altaffilmark{3},
Nate~B.~Lust\altaffilmark{1,4},
Jasmina~Blecic\altaffilmark{1,5}, and
Madison~Stemm\altaffilmark{1}
}

\affil{\sp{1} Planetary Sciences Group, Department of Physics,
              University of Central Florida, Orlando, FL 32816-2385 \\
       \sp{2} Space Research Institute, Austrian Academy of Sciences,
              Schmiedlstrasse 6, A-8042 Graz, Austria \\
       \sp{3} Cornell Center for Astrophysics and Planetary Sciences,
              Space Sciences Building, Cornell University, Ithaca,
              NY 14853-6801 \\
       \sp{4} Department of Astrophysical Sciences, Princeton University,
              Princeton, NJ 08544 \\
       \sp{5} Department of Physics, New York University Abu Dhabi,
              PO Box 129188 Abu Dhabi, UAE}
\email{patricio.cubillos@oeaw.ac.at}

\begin{abstract}

  Time-correlated noise is a significant source of uncertainty when
  modeling exoplanet light-curve data.  A correct assessment of
  correlated noise is fundamental to determine the true statistical
  significance of our findings.  Here we review three of the most
  widely used correlated-noise estimators in the exoplanet field, the
  time-averaging, residual-permutation, and wavelet-likelihood
  methods.
  We argue that the residual-permutation method is unsound 
  in estimating the uncertainty of parameter estimates. We thus
  recommend to refrain from this method altogether.  We characterize
  the behavior of the time averaging's rms-vs.-bin-size curves
  at bin sizes similar to the total observation
  duration, which may lead to underestimated uncertainties.  For the
  wavelet-likelihood method, we note errors in the published equations
  and provide a list of corrections.  We further assess the
  performance of these techniques by injecting and retrieving
  eclipse signals into synthetic and real Spitzer light
  curves, analyzing the results in terms of the
  relative-accuracy and coverage-fraction statistics.  Both the
  time-averaging and wavelet-likelihood methods significantly improve
  the estimate of the eclipse depth over a white-noise analysis (a
  Markov-chain Monte Carlo exploration assuming uncorrelated
  noise).  However, the corrections are not perfect, when retrieving
  the eclipse depth from Spitzer datasets, these methods covered the
  true (injected) depth within the 68\% credible region in only
  $\sim$45--65\% of the trials.  Lastly, we present our open-source
  model-fitting tool, Multi-Core Markov-Chain Monte Carlo ({MC$^3$}).
  This package uses Bayesian statistics to estimate the best-fitting
  values and the credible regions for the parameters for a
  (user-provided) model.  {MC$^3$} is a Python/C code, available
  at \href{https://github.com/pcubillos/MCcubed}
  {https://github.com/pcubillos/MCcubed}.

\end{abstract}
\keywords{methods: statistical ---
          planets and satellites: fundamental parameters ---
          techniques: photometric
}

\section{INTRODUCTION}
\label{sec:introduction}

Whether one's goal is the detection or the characterization of
exoplanets through transit or eclipse observations, the large
contrast between the stellar and planetary emission (e.g., about a thousand
times in the infrared for a hot Jupiter around an FGK dwarf
  star) make the data analysis an intrinsically
challenging task.  For example, for the Spitzer Space Telescope, most
planetary signals \citep[e.g.,][]{StevensonEtal2010natGJ436b,
 DemoryEtal2012apjl55cnceEclipse} lie below the instrument's design
 criteria for
photometric stability \citep{FazioEtal2004apjsIRAC}.
Extracting planetary signals at this precision requires meticulous
data reduction.  Despite our best attempts to account for all known
systematics, time-correlated residuals (or red noise)
between the data and models often remain.  These systematics may originate
from instrumental or astrophysical sources, for example: stellar flux
variations from flares or granulation; imperfect flat fielding; or
telluric variations from 
changing weather conditions, differential extinction, or imperfect
telescope systematics corrections from changing telescope pointing.
Many authors have acknowledged correlated
noise as an important source of noise in time-series datasets
\citep[e.g.,][]{PontEtal2006mnrasRednoise, WinnEtal2007ajHATP1b,
 AgolEtal2010apjHD189, CubillosEtal2013apjWASP8b}.

Correlated noise affects both the accuracy and the precision of
estimates of the model parameters.  The typical statistical analyses
neglect the correlation between data points (e.g., likelihood
functions such as $\chi\sp{2}$, based on uncorrelated noise).  Hence,
their estimated best-fitting values may be biased, whereas their
credible regions (Appendix \ref{sec:CredRegion}) can be incorrect.
This paper reports our study of three common correlated-noise
estimators found in the exoplanet literature; specifically, in
analyses of transit and eclipse exoplanet light curves.  First, the
time-averaging method \citep{PontEtal2006mnrasRednoise,
WinnEtal2007ajHATP1b} compares the standard deviation of the data to
the (expected) uncorrelated-noise standard deviation, scaling the
uncertainties accordingly.  Next, the residual-permutation (or
``prayer bead'') method \citep{BouchyEtal2005CorrNoise} uses a
data-shifting (bootstrap-like) algorithm that preserves the structure
of the residuals.  Lastly, the method
of \citet[][]{CarterWinn2009apjWavelets} calculates the likelihood
function in a wavelet basis, where the correlation between the wavelet
coefficients is negligible.  Qualitatively speaking, these methods do
return larger parameter uncertainties for stronger correlated noise.
However, besides \citet{CarterWinn2009apjWavelets}, there are few
efforts to validate their quantitative accuracy.  We have implemented
these methods, testing them with real and synthetic exoplanet eclipse
data.

Although recently there has been development of additional sophisticated
methods to model exoplanet light
curves \citep[e.g.,][]{GibsonEtal2012mnrasGaussProc,
JordanEtal2013apjWASP6bPCA, Morello2015apjPixelICA,
EvansEtal2015mnrasHD209bGaussianProcesses}, we explicitly exclude
them from this study because the large number of runs and dataset
sizes would require unfeasible amounts of computing power.  For
example, Gaussian processes can become computationally prohibitive for
datasets larger than $\sim 1000$ data
points \citep[][]{Gibson2014mnrasModelSelecGP}.

With a focus on atmospheric characterization, we concentrate on
estimating the eclipse depths from Spitzer exoplanet light curves,
since they represent the largest and best-quality sample of exoplanet
data beyond 2 {\microns}.  Spitzer data are affected by two well-known
systematics: time-varying sensitivity (ramp) and intra-pixel
sensitivity variations \citep{KnutsonEtal2009apjHD149026bphase,
CharbonneauEtal2005apjTrES1}.  Although several models have been
proposed to correct for these systematics
\citep{HarringtonEtal2007natHD149026b, KnutsonEtal2008apjHD209,
BallardEtal2010paspIntraPixel, AgolEtal2010apjHD189,
 StevensonEtal2012apjHD149026b, LewisEtal2013apjHAT2bPhase,
 DemingEtal2014IntraPixelModel}, the corrections are not always
 perfect, and thus many light-curve fits exhibit time-correlated
 residuals.

In Section \ref{sec:impact}, we discuss the impact of correlated noise
on determining model-parameter uncertainties.  In
Section \ref{sec:estimators}, we review the most commonly used
correlated-noise estimators in exoplanet analyses.  In
Section \ref{sec:simulations}, we test and compare the
correlated-noise methods by retrieving synthetic eclipse curves that
were injected into synthetic and real light-curve data.  In
Section \ref{sec:mc3} we present our open-source package, Multi-Core
Markov-Chain Monte Carlo (MC\sp{3}), to calculate the
model-parameters' credible regions.  Finally, in Section
\ref{sec:conclusions} we present our conclusions.

\section{THE IMPACT OF CORRELATED NOISE}
\label{sec:impact}

A central ingredient in both frequentist and Bayesian parametric
modeling is the sampling distribution for the data: the joint
probability density function (PDF) for the data values, as a function
of the model parameters.  In our time series setting, we denote the
data by $\yvec = (y_1,\ldots,y_n)$, with $y_i$ denoting the value of a
measurement at time $t_i$.  Similarly, we denote the model predictions
by $\fvec = (f_1,\ldots,f_n)$, with $f_i = f(t_i;\params)$ for a model
function with parameters $\params$.  Note that the predictions are
functions of the parameters, $f_i(\params)$, but we often suppress the
parameter dependence for convenience.  The sampling distribution is
the $n$-dimensional joint PDF, $p(\yvec|\params)$.

When $\yvec$ is fixed to an actually observed data vector, the
sampling distribution as a function of the model parameters is called
the likelihood function, $\like(\params)$.
Bayesian methods quantify uncertainty in the parameters via the
dependence of $\like(\params)$ on the parameters; Bayes's theorem and
the law of total probability convert this dependence into posterior
probabilities for statements about the parameters.
Frequentist methods quantify uncertainty by first defining statistics
(functions of $\yvec$) that produce point estimates or intervals in
the parameter space (perhaps using the $\params$ dependence of the
likelihood function), and then using the sampling distribution to
quantify the variability of the statistics across ensembles of
hypothetical data vectors.
The variability in the sample space then is mapped into uncertainty
quantifications in the parameter space (e.g., bias of a point
estimate, or coverage of a confidence interval).

Commonly, the data are modeled as the sum of the predictions and
independent, zero-mean, normally-distributed noise,
\begin{equation}
y_i = f_i(\params) + \eps_i,
\label{y-fe}
\end{equation}
with independent noise probabilities
\begin{equation}
p(\eps_i) = \frac{1}{\sigma_i\sqrt{2\pi}}\exp\left[-\frac{\eps_i^2}{2\sigma_i^2}\right],
\label{e-norm}
\end{equation}
where $\sigma_i$ is the standard deviation for the noise contribution
in measurement $i$.  In this scenario, the sampling distribution
factors,
\begin{align}
p(\yvec|\params)
  & = \prod_i p(y_i|\params) \nonumber\\
  & =\frac{1}{{(2\pi)}^{n/2}}
      \left(\prod_i\sigma_i\right)^{-1} 
      \exp\left[-\frac{1}{2}\sum\sb{i} \frac{\eps\sb{i}(\params)\sp{2}}{\sigma\sb{i}\sp{2}}\right],
\label{eq:samp-ind}
\end{align}
where the sum inside the exponential is the familiar ``$\chi^2$'' sum
of squared, standardized residuals.  Parameter estimation based on
maximizing Eq.\ (\ref{eq:samp-ind}) with respect to $\theta$ is
called \emph{weighted least squares} (WLS) regression when
$f_i(\theta)$ is linear with respect to the parameters.  If in
addition the $\sigma_i$ values are all the same (homoskedastic), the
approach is called \emph{ordinary least squares} (OLS).

In the general case, correlated Gaussian noise has a multivariate
normal PDF with a non-diagonal precision matrix $K$,
\begin{equation}
p(\eps) = \frac{\vert K\vert^{1/2}}{(2\pi)^{n/2}}
  \exp\left[-\frac{1}{2}\sum_{ij} \eps_i K_{ij} \eps_j\right],
\label{eq:corrPDF}
\end{equation}
where $\vert K\vert$ denotes the determinant of the precision matrix.
The derivation to the specific case of uncorrelated Gaussian noise is
trivial.  As a consequence of independence, $K$ becomes a diagonal
matrix with components
\begin{align}
K_{ij} = \frac{1}{\sigma_i\sigma_j} \delta_{ij}.
\end{align}
Therefore, the sum in Eq.\ (\ref{eq:corrPDF}) collapses to the single
sum in Eq.\ (\ref{eq:samp-ind}).

Parameter estimation based on maximizing this likelihood
function with respect to $\theta$ is called \emph{generalized least
squares} (GLS) regression when $f_i(\theta)$ is linear with respect to
the parameters.  When the model has nonlinear dependence on any of the
parameters, maximum likelihood estimate is the \emph{nonlinear least
squares} estimate, regardless of whether the errors are correlated or
homoskedastic.

Methods currently used to handle correlated noise fall into two broad
classes.  Methods like the \citet{CarterWinn2009apjWavelets} wavelet
approach estimate a correlation matrix, and produce GLS estimates.  In
contrast, the time-averaging and residual-permutation approaches rely
on WLS for estimation, but devise rules for inflating uncertainties to
account for the ignored correlations.  It is well known that
non-linear least squares estimates are statistically consistent; that
is, asymptotically (as $n\rightarrow\infty$), estimates converge to
the true parameter values \citep{Wu1981NonlinearLSE}.  Roughly
speaking, although correlation complicates the way information
accumulates across samples, infinite sample size ameliorates the
complications.  However, since WLS does not account for correlations,
the quality of estimates can be significantly compromised with finite
sample size.

In Appendix \ref{sec:OLSGLSexamples} we describe two simple example
calculations comparing WLS and GLS that provide some
insight into the costs of ignoring noise correlation.  The first
example treats estimation of the amplitude of a \emph{constant signal}
in the presence of autoregressive noise, a simple and analytically
tractable example of correlated noise.  When noise is independent,
with standard deviation $s$, the uncertainty in an estimate of a
constant signal level is $s/\sqrt{n}$, the familiar ``root-$n$'' law.
The WLS estimate has
just this behavior.  In contrast, when the noise is known to be
positively correlated between adjacent samples, the uncertainty in the
GLS estimate decreases more slowly
than $1/\sqrt{n}$.  This kind of example motivates approaches like
time averaging that attempt to account for correlation merely by
inflating uncertainties.

The second example replaces the constant signal with an eclipse-like
\emph{dip signal}.  The dip location and width are presumed known;
the background level and dip depth are to be estimated.  In this case,
a simple simulation study shows that noise correlation does not merely
inflate uncertainties.  It can also corrupt parameter estimates, with
WLS estimates potentially taking values far away from the optimal
estimates that account for noise correlations.  This occurs when
parameters of interest pertain to temporally localized structure in
the model, for which noise correlations can significantly change the
data projections needed for accurate inference.

Together, these examples show that methods that seek to account for
correlations only by inflating parameter uncertainties are at best
suboptimal (producing larger estimation errors than could be achieved
with a good correlated noise model), and can sometimes be
significantly misleading.

\section{COMPUTING PARAMETER UNCERTAINTIES}
\label{sec:estimators}

\subsection{Markov-chain Monte Carlo}
In the Bayesian framework, a credible
region for the parameters of a model, $\mathcal{M}$, can
be computed via the Markov-chain Monte Carlo
(MCMC) algorithm.  The MCMC method generates a large number of random
samples from the parameter  space with a probability
density proportional to the posterior probability distribution:
\begin{equation}
p(\theta|\yvec, \mathcal{M}) \propto p(\theta|\mathcal{M})
                               p(\yvec|\theta, \mathcal{M}),
\end{equation}
where $p(\theta|\mathcal{M})$ is the prior probability distribution.
A marginal highest-posterior-density (HPD) credible region for each
parameter is then obtained from the interval that contains a certain
fraction of the highest posterior density (typically 68\%, 95\%, or
99\%) of the marginalized posterior (see
Appendix \ref{sec:CredRegion}).  For example, when the posterior
follows a normal distribution, the 68.3\% marginal credible interval
corresponds to the interval contained within one standard deviation
from the mean.

Inference based on the likelihood function of Eq.\ (\ref{eq:samp-ind})
works well when the noise contributions are independent and normally
distributed; however, it does not account for time-correlated noise.
Alternatively, an inference that uses the full covariance matrix, as
in of Eq.\ (\ref{eq:corrPDF}), should account for correlated noise,
although its calculation often becomes computationally prohibitive.

\subsection{Time Averaging}
\label{sec:timeavg}

\citet{PontEtal2006mnrasRednoise} developed a method
to compute the uncertainty of a transit or
eclipse-depth estimation using the
light-curve data points themselves.  They considered the noise as the
sum in quadrature of two components, a purely white (uncorrelated)
source (characterized by a standard deviation per data point
$\sigma\sb{w}$), and a purely time-correlated source (characterized by
$\sigma\sb{r}$).
\citet{PontEtal2006mnrasRednoise} assumed the white-noise
component to scale as $\sigma\sb{w}/\sqrt{n}$, with $n$ the number of
data points in the transit; whereas the time-correlated standard
deviation, $\sigma\sb{r}$, to be independent of the number of data
points. Then
for any given signal, the
uncertainty of a measurement should scale as:
\begin{equation}
\sigma\sb{d} = \sqrt{\frac{\sigma\sb{w}\sp{2}}{n} + \sigma\sb{r}\sp{2}}.
\label{eq:deptherror}
\end{equation}

For small $n$, $\sigma\sb{d}$ may be dominated by
$\sigma\sb{w}/\sqrt{n}$, whereas as $n$ increases, $\sigma\sb{d}$
approaches $\sigma\sb{r}$.  The time-averaging method uses this fact
to estimate the contribution from the correlated noise.
Note, however, that this hypothesized behavior is not typical of
stationary correlated noise models exhibiting long-range dependence,
which instead have $\sigma\sb{d}$ decreasing at a rate slower than
$1/\sqrt{n}$, but still monotonically decreasing to
zero \citep{BeranEtal2013LongMemoryProcesses}.

We implement the time-averaging procedure as described
by \citet{WinnEtal2007ajHATP1b}.  First, we calculate the residuals
between the data points and the best-fitting model.  Then, we group
the residuals in time-ordered, non-overlapping bins of $N$ elements
each, and calculate their mean values.  Lastly, we calculate the
standard deviation (or root mean squared, rms) of the binned
residuals, rms$\sb{N}$.  We repeat the process for a range of bin
sizes from one to half the data size.  The uncertainty of rms$\sb{N}$
is approximately $\sigma\sb{\rm rms}=$\,${\rm rms}\sb{N}/\sqrt{2M}$
(see Appendix \ref{sec:StdUncert}).

Now, let $\sigma\sb{1}$ be the rms value of the non-binned residuals
(presumed to be dominated by white noise).  In the absence of
correlated noise, the expected rms for the set of $M$ bins,
each containing $N$ points, is given by the extrapolation of
$\sigma\sb{1}$
\citep{WinnEtal2008apjXO3bRedNoise}:
\begin{equation}
\sigma\sb{N} = \frac{\sigma\sb{1}}{\sqrt{N}} \sqrt{\frac{M}{M-1}}.
\label{eq:rmsvsbin}
\end{equation}

\begin{figure}[tb]
\centering
\includegraphics[width=\linewidth, clip]{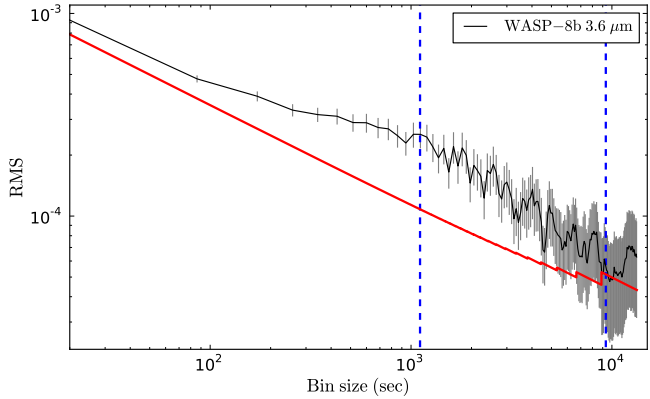}
\caption{Binned residuals rms vs.\ bin size (black curve with gray
 error bars) of WASP-8b Spitzer eclipse at 3.6 {\microns} (PI
 J. Harrington, Program ID 60003,
 see \citealp{CubillosEtal2013apjWASP8b}).  The red curve corresponds
 to the expected rms for white noise (Equation \ref{eq:rmsvsbin}).
 The saw-tooth look of the curve arises from the discreet change
 in $M$, which becomes more significant as $N$ increases.
 The vertical dashed lines mark the duration of ingress/egress (left)
 and eclipse (right).  The gray vertical error bars denote the
 $1\sigma$ uncertainty of the rms residuals (${\rm rms}\sb{N}/\sqrt{2M}$).}
\label{fig:wasp8bRMS}
\end{figure}

The rms$\sb{N}$ and $\sigma\sb{N}$ curves are analogous to
$\sigma\sb{d}$ and $\sigma\sb{w}$, respectively.  The time-averaging
correction inflates the data uncertainties multiplying them by the
ratio $\beta\sb{N} = {\rm rms}\sb{N} / \sigma\sb{N}$ if
$\beta\sb{N}$ is statistically larger than one (i.e., by more than
$1\sigma\sb\beta = \sigma\sb{\rm rms}/\sigma\sb{N}$).  Finally, one
runs a white-noise MCMC analysis, i.e., with Eq.\ (\ref{eq:samp-ind}),
using the inflated data errors.  One typically visualizes both curves
in an rms vs.\ bin size plot (Figure \ref{fig:wasp8bRMS}).

\subsubsection{Behavior at Large Bin Sizes}
\label{sec:TAzerotest}

One has to be aware that the binned-rms uncertainty $\sigma\sb{\rm
rms}$ is an asymptotic approximation.  In the large bin-size regime
(equivalently, small $M$) that approximation is not justified. The
reason is that the marginalized posterior distribution for ${\rm
rms}\sb{N}$ (Eq. \ref{eq:fSigma}), which has the form of an
inverse-gamma distribution, becomes increasingly skewed as $M$
decreases.

By comparing the 68\%-credible-region error bars
of the inverse-gamma formula with the asymptotic approximation, 
we find that the latter moderately
overestimates the lower error bar by 5\%--10\%
for $M<200$.  For the upper error bar the asymptotic approximation
underestimates the error by 5\% to 450\% between $M=200$ and $M=2$.
If one does not consider the skewed error bars, it may seem that the
rms curves deviate below the $\sigma\sb{N}$ curve at large bin
sizes \citep[e.g.,][]{StevensonEtal2012apjHD149026b,
CubillosEtal2013apjWASP8b, BlecicEtal2013apjWASP14b}. However, this
deviation is not statistically significant when one computes the
uncertainties with the correct posterior distribution.

Since the typical transit (or eclipse) observation does not last much
longer than the duration of the transit itself (usually one--two hours
of out-of-transit before and after), it is important to
consider the asymmetric rms error bars to properly account for
correlated noise.  Furthermore, since the signal-to-noise ratio of
rms$_N$ decreases proportionally to $\sqrt{M}$ as $N$ increases, one
ideally wants the longest-possible observation duration to improve the
constraint on $\beta\sb{N}$ at the desired timescale of the event.

\subsection{Residual Permutation}
\label{sec:prayer}

Residual permutation (also called the prayer bead method) is inspired
by nonparametric bootstrapping methods from frequentist statistics.
Nonparametric bootstrap methods directly use the sampled data
(typically via resampling) to generate a distribution that
approximates the sampling distribution, $p(\yvec| \theta^*)$, for the
true parameter values, $\theta^*$.  Nonparametric bootstrapping
typically relies on independent resampling of the data or residuals
(possibly re-scaled), with
replacement \citep[e.g.,][]{DavisonHinkley1997BootstrapMethods,
RuppertMatteson2015StatisticsForFinancialEngineering}.

The motivating idea of the residual-permutation approach is to shift
the data while preserving the time ordering and, thus, preserving the
correlation structure.  While the structure is indeed preserved, the
residual-permutation method does not resample with replacement, a
crucial requrement for nonparametric bootstrapping to work, i.e., to
produce independent replicated datasets (a resample of the entire
time-series observation).  When there is correlated noise, the shifted
datasets do not correspond to an independent resampling from any
distribution, and thus do not exhibit the variability necessary for
correct uncertainty quantification (e.g., computing confidence levels
or estimator bias).

In the exoplanet field, the residual-permutation technique has been
repeatedly used to estimate parameter uncertainties.  However, the
name of the technique has been loosely used to describe similar, but
not equivalent procedures over the past decade.
\citet{BouchyEtal2005CorrNoise}, \citet{GillonEtal2007aaGJ436bspitz},
and \citet{Southworth2008HomogeneousStudyI} all describe different
methods, when referring to residual permutations.  Some authors
reference \citet{JenkinsEtal2002apjDetectionConfidence}, who actually
use a ``segmented bootstrap'', applying the method for detection
instead of parameter estimation.  Furthermore, several authors have
wrongly attributed the method
to \citet{MoutouEtal2004aaOgle132b}. Thus, there is a visible lack of
rigor in the use of this method.

Currently, the most widely-used version of residual permutation is the
one described by \citet{Southworth2008HomogeneousStudyI} or
\citet{WinnEtal2008apjXO3bRedNoise}.  This implementation computes the
residuals between the light curve and the best-fitting model,
cyclically shifts the residuals (preserving the point-to-point
structure and thus the ``redness'' of the noise) by a given number of
data points, adds the residuals back to the model, and finds a new set
of best-fitting parameters.  Usually, either one repeats the
shift--fit process for a large number of iterations with random
shifts, or one sequentially shifts the residuals by one data point at
a time, fitting all possible shifts.  Each parameter uncertainty is
then given by the respective standard deviation of the distribution of
best-fitting values.  As already noted, this does not correspond to a
sound resampling procedure, thus we will not consider residual
permutations for the subsequent analyses.

There is a significant literature on generalizing the independent and
identically distributed nonparametric bootstrap idea to address time
series problems with correlated noise; this is a topic of ongoing
research.  One widely used approach is the block bootstrap.  Presuming
the investigator knows or can estimate a longest scale for
correlations, $\Delta t$, the data are divided into blocks of length
greater than $\Delta t$, and bootstrap resampling is done by drawing
blocks of data at random to build a replication.  A particular block
rigidly preserves the time ordering of a subset of the data; in
replications, it will appear shifted in time by various amounts.  This
behavior resembles the behavior of the prayer bead method.  But block
resampling produces greater variability than shifting the entire data
vector, and by sampling with replacement, it produces ensembles that
approximate independent draws from a (dependent) sampling
distribution.  The ``segmented bootstrap'' devised
by \citet{JenkinsEtal2002apjDetectionConfidence} for analysis of
ground-based transit photometry is similar to the block bootstrap.
The block bootstrap only works if the correlation scale is
significantly shorter than the span of the data, which will often not
be true for Spitzer exoplanet eclipse data, so we do not consider it
further here.  Further details about the block bootstrap and other
methods for resampling dependent data may be found
in \citet{Lahiri2003}.

\subsection{Wavelet Analysis}
\label{sec:wavelet}

\citet{CarterWinn2009apjWavelets} introduced to the exoplanet field a
technique where the time-correlated noise is modeled using wavelet
transforms \citep{DericheTewfik1993tspLikeFractalNoise,
Wornell1993ieeeWavelet,
WornellOppenheim1992ieeeFractalSignalEstimation,
WornellOppenheim1992ieeeWaveletFractalModulation}.  This method
projects the time series residuals into an orthonormal wavelet basis,
where the off-diagonal terms of the covariance matrix become
negligible, thus simplifying the likelihood function calculation.
Furthermore, they assumed noise that has a power spectral density with
frequency $f$, varying as $1/f\sp{\gamma}$.  They parameterized the
noise with three parameters, $\gamma$, $\sigma\sb{\omega}$, and
$\sigma\sb{r}$, as described in Equations (41)--(43)
of \citet{CarterWinn2009apjWavelets}.

A thorough review of wavelets is beyond the scope of this work;
see \citet{Mallat2008WaveletTour} and \citet{Wornell1996Wavelet} for
more comprehensive discussions.  Briefly, a wavelet transform projects
a time-series signal onto a basis of functions that are dilations and
translations of a compact parent (``wavelet'') function.  The
resulting transform has two dimensions, scale and location (in time).
The discrete wavelet transform (DWT) consists of the hierarchical
application over $M$ dilation scales of an orthonormal wavelet
transform on a discrete time-series signal.  For a signal consisting
of $N=N\sb{0}2\sp{M}$ uniformly-spaced samples (with $N\sb{0}$
integer), and a wavelet function with $2N\sb{0}$ coefficients, the DWT
produces $N\sb{0}$ scaling coefficients and $N\sb{0}2\sp{m-1}$ wavelet
coefficients at each scale $m$, totaling $N\sb{0}(2\sp{M}-1)$ wavelet
coefficients.

\citet{CarterWinn2009apjWavelets} recommend the fourth-order
Daubechies wavelet basis \citep{Daubechies1988wavelets} for modeling
time-series correlated noise, which we adopt in the current work.
This is a basis well localized in time and frequency, well suited for
$1/f\sp{\gamma}$ noise \citep{Wornell1996Wavelet}.
\citet{CarterWinn2009apjWavelets} found that correlations between the
wavelet and scaling coefficients decays faster for the Daubechies
basis than the Haar basis, producing negligible covariances.  Another
advantage is that since the Daubechies basis is well localized in
time, it reduces artifacts arising from the assumption of a periodic
boundary condition by the wavelet transform.

\subsubsection{Wavelet-based Likelihood}
\label{sec:waveletlike}

The likelihood function in the wavelet analysis is calculated in the
following way.  Let $\epsilon(t)$ be the fitting residuals of a
time-series signal.  Considering $\epsilon(t)$ as the contribution of
a time-correlated ($\gamma \neq 0$) and an uncorrelated ($\gamma = 0$)
component:
\begin{equation}
\epsilon(t) = \epsilon\sb{\gamma}(t) + \epsilon\sb{0}(t),
\end{equation}
this method calculates the DWT of $\epsilon(t)$ to produce the
wavelet, $r\sb{n}\sp{m}$, and scaling, $\bar{r}\sb{n}\sp{1}$,
coefficients of the signal.  The variances of these coefficients are
computed, respectively, as:
\begin{eqnarray}
\label{eq:varwavelet}
\sigma\sb{W}\sp{2} & = & \sigma\sb{r}\sp{2}2\sp{-\gamma m}          +
                        \sigma\sb{\omega}\sp{2} \\ 
\label{eq:varscale}
\sigma\sb{S}\sp{2} & = & \sigma\sb{r}\sp{2}2\sp{-\gamma}  g(\gamma) +
                        \sigma\sb{\omega}\sp{2},
\end{eqnarray}
where $\sigma\sb{\omega}$ and $\sigma\sb{r}$ parameterize the standard
deviation of the uncorrelated and the correlated-noise signals,
respectively, and $g(\gamma)=1/(2^{1-\gamma}-1)$ for
$\gamma\ne1$ \citep[following derivations from, e.g.,
][]{FadiliBullmore2002, Wornell1993ieeeWavelet} and $g(\gamma) =
1/2\ln2$ for $\gamma=1$ \citep{CarterWinn2009apjWavelets} (see
Appendix \ref{sec:WaveletVariance}).  Therefore, the wavelet-based
likelihood function is given by
\begin{eqnarray}
\mathcal{L}({\bf x}, \sigma\sb{\omega}, \sigma\sb{r}) 
       & = & \left\{ \prod\sb{m=1}\sp{M} \prod\sb{n=1}\sp{N\sb{0}2\sp{m-1}}
             \frac{1}{\sqrt{2\pi\sigma\sb{W}\sp{2}}}
             \exp\left[{-\frac{(r\sb{n}\sp{m})\sp{2}}{2\sigma\sb{W}\sp{2}}} \right] \right\} \times \nonumber \\
         &  & \left\{ \prod\sb{n=1}\sp{n\sb{0}}
              \frac{1}{\sqrt{2\pi\sigma\sb{S}\sp{2}}}
              \exp\left[{-\frac{(\bar{r}\sb{n}\sp{1})\sp{2}}{2\sigma\sb{S}\sp{2}}} \right] \right\}.
\label{eq:wavelike}
\end{eqnarray}

Equation (\ref{eq:wavelike}) allows one to fit a model, sample its
parameter's posterior distribution, and determine the credible
regions, while taking into account the effects of time-correlated
noise.

During our review and implementation of the wavelet-likelihood
technique from \citet{CarterWinn2009apjWavelets}, we found a few
oversights in their equations and code (available in the Astronomical
Source Code Library,
ASCL\footnote{\href{http://asterisk.apod.com/viewtopic.php?f=35\&t=21675}
{http://asterisk.apod.com/viewtopic.php?f=35\&t=21675}}). See details
in Appendix \ref{sec:errata}.

\begin{figure*}[tb]
\centering
\includegraphics[width=\linewidth, clip]{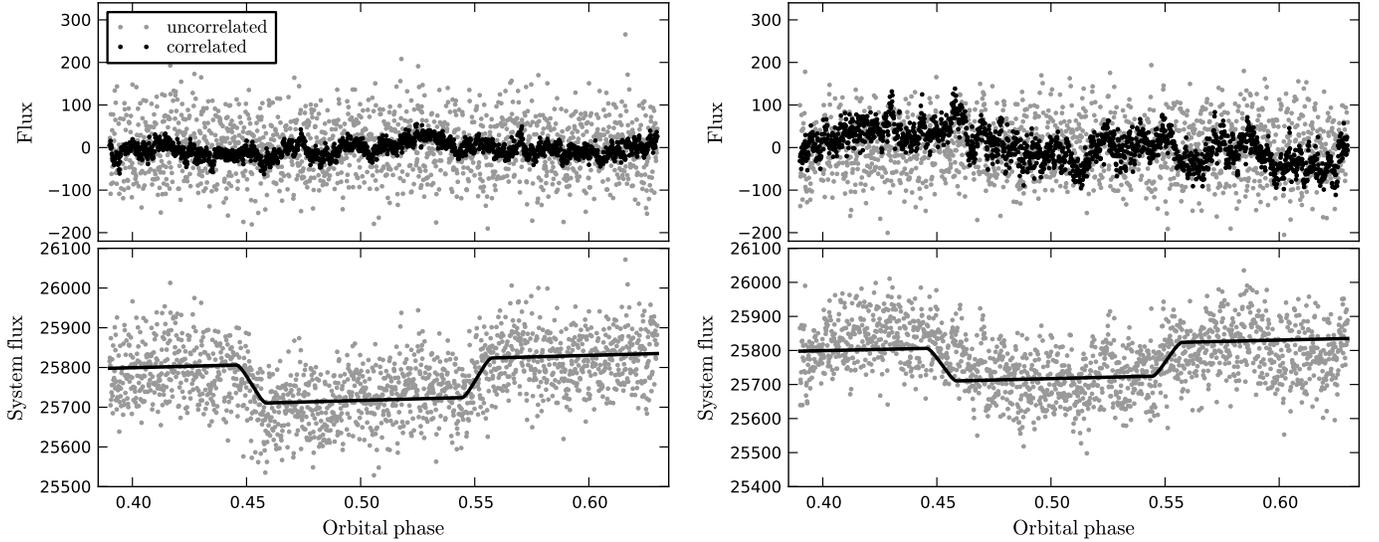}
\caption{Simulated Spitzer time-series datasets.  The top panels
 overplot the correlated (black) and uncorrelated (grey) noise
 components of the light curves vs.\ orbital phase.  The bottom panels
 show the synthetic light curves (eclipse, ramp, and noise)
 vs.\ orbital phase in gray.  The black solid line shows the noiseless
 model.  The noise rms ratios are $\alpha=0.25$ (left panels) and
 $\alpha=0.5$ (right panels).}
\label{fig:synthetic}
\end{figure*}

\section{CORRELATED-NOISE TESTS FOR EXOPLANET ECLIPSE DATA}
\label{sec:simulations}

To assess the performance of the correlated-noise estimators described
in Section \ref{sec:estimators}, we carried out injection--retrieval
eclipse simulations.  We focus on estimating the secondary-eclipse
depth in a light curve observation, creating synthetic light curves
that resemble Spitzer InfraRed Array Camera (IRAC) observations in
terms of the signal-to-noise ratio (S/N), known systematics, cadence,
observation duration, and eclipse shape.

In our first experiment we test the estimators' performances when the
time-correlated noise is described by a stochastic wavelet signal with
a $1/f$ power spectral density \citep[similar to the experiment
of][]{CarterWinn2009apjWavelets}.  We test the case when the
observation time span is similar to the eclipse-event duration
(Section \ref{sec:synthetic}, typical of real Spitzer secondary
eclipse observations) and for the hypothetical case when the time span
lasted an order of magnitude longer than the eclipse event (Section
\ref{sec:synthetic2}).  In a second experiment (Section
\ref{sec:semisynth3}) we test the estimators on a more realistic
case by injecting a synthetic eclipse signal into Spitzer
phase-curve datasets.

\subsection{Synthetic-noise Simulation}
\label{sec:synthetic}

In this simulation we generate synthetic light curves by combining
a \citet{MandelAgol2002ApJtransits} eclipse model, a linear ramp
model, and a signal with both correlated and uncorrelated noise.  The
light-curve parameters closely follow those of a Spitzer observation
of the WASP-12 system (Table \ref{table:parameters}).  The signal
consists of 1700 data points, with a cadence of $\sim$12 seconds
between data points, spanning an orbital-phase range from 0.39 to
0.63, about twice the eclipse duration.

\begin{table}[ht]
\centering
\caption{\label{table:parameters} Synthetic light curve parameters}
\begin{tabular}{lc}
\hline
\hline
Parameter                           & Value           \\
\hline
Eclipse depth (counts)              & 98.1            \\
Eclipse duration (phase)            & 0.1119          \\
Eclipse mid point (phase)           & 0.5015          \\
Eclipse ingress/egress time (phase) & 0.013           \\
Ramp slope (counts/phase)           & 0.006           \\
System flux (counts)                & 25815           \\
$\sigma\sb{\omega}$ (counts)        & 64.5            \\
$\sigma\sb{r}$ (counts)             & 0, 230, and 459 \\
\hline
\end{tabular}
\end{table}

We created three sets of 5000 light-curve realizations each.  For each
realization, we generate a zero-mean random normal distribution, which
we add to the light curve as the uncorrelated noise signal.  We adjust
the variance of this signal ($\sigma\sb{\omega}\sp{2}$) to yield an
eclipse-depth signal-to-noise ratio of 30.  Additionally, we generate
purely-correlated $1/f$ signals ($\sigma\sb{\omega}=0$) using a
Gaussian random number generator to produce wavelet coefficients with
variances given by Equations (\ref{eq:varwavelet}) and
(\ref{eq:varscale}).  Then, we apply the inverse DWT to transform the
signal from the wavelet basis to the time domain.  Following the
notation of \citet{CarterWinn2009apjWavelets}, we denote by $\alpha$
the ratio between the rms of the uncorrelated and correlated noise
signals.

We constructed the signals in each of the three sets to have a pure
uncorrelated noise, a weak time-correlated signal, and a strong
correlated signal ($\alpha=0.0$, $0.25$, and $0.5$, respectively).
Figure \ref{fig:synthetic} shows two synthetic light curves for
$\alpha=0.25$ and $0.5$.  Note that our designations of ``weak'' and
``strong'' are, to some extent, arbitrary.  We selected these limits
based on our experience and tests: for $\alpha \lesssim 0.20$, the
time-correlated signal becomes negligible compared to the
uncorrelated-noise signal, whereas values of $\alpha \sim 0.5$ are on
the level of what we have observed in some cases \citep[e.g.,
WASP-8b,][]{CubillosEtal2013apjWASP8b}.

For each realization, we compute the parameter posteriors using the
methods described in Section \ref{sec:estimators}, excluding residual
permutation, which we deem to be unsound.  Our model-fitting routines
only fix the eclipse ingress/egress-time parameter (usually
poorly-constrained by eclipse data), leaving free the system flux,
eclipse depth, eclipse midpoint, eclipse duration, and ramp slope.
First, we carry out a ``white analysis'' (i.e., ignoring the
time-correlation between data points) by using Equation
(\ref{eq:samp-ind}) to compute the model-parameter best-fitting values
(using the Levenberg-Marquardt algorithm) and their posterior
distributions (using a MCMC).

\begin{figure}[tb]
\centering
\includegraphics[width=\linewidth, clip]{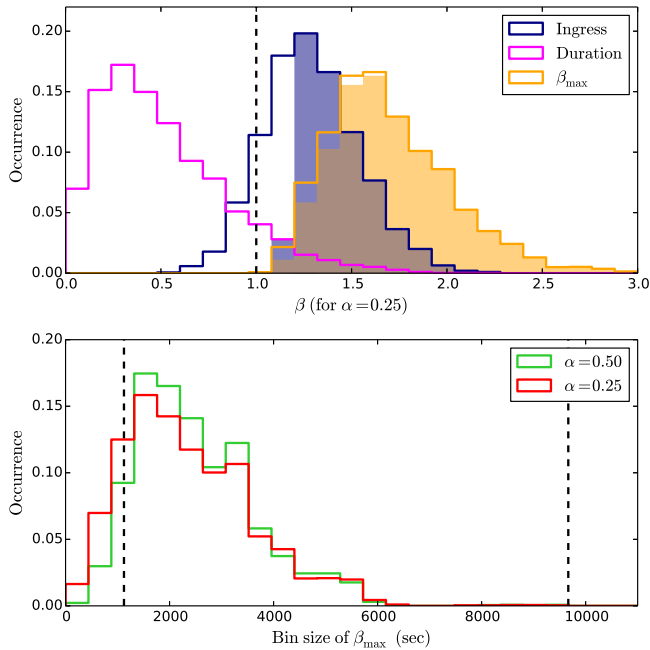}
\caption{{\bf Top:} Normalized distribution of $\beta$ for the
 $\alpha=0.25$ set.  The histograms represent $\beta$ measured at the
 ingress-time and eclipse-duration timescales, and at the maximum
 value of $\beta$.  The vertical dashed line marks $\beta=1$. The
 colored shaded areas denote the corresponding fraction of trials that
 were more than $1\sigma\sb{\beta}$ greater than one. The
 distributions for the other two sets ($\alpha=0$ and 0.5) were
 similar.  {\bf Bottom:} Normalized distribution of the bin sizes for
 $\beta\sb{\rm max}$.  The vertical dashed lines indicate the ingress
 time and eclipse duration.}
\label{fig:beta}
\end{figure}

Next, we use the best-fit results to calculate the time-averaging
rms-vs.-bin size curves.  We retrieve the $\beta$ factor at three
timescales: at the ingress time, at the eclipse duration, and at the
time of maximum $\beta$ ($\beta\sb{\rm max}$, Figure \ref{fig:beta}).
In accordance with the discussion in Section \ref{sec:TAzerotest},
most $\beta$ values at the eclipse-duration timescale (similar to the
total observation duration) were not significant. Thus, we adopted
$\beta\sb{\rm max}$ as the scaling factor to calculate the
time-averaging method uncertainties.  Finally, we apply the
wavelet-based likelihood method in an MCMC guided by Equation
(\ref{eq:wavelike}), simultaneously fitting the noise parameters
($\sigma\sb{\omega}$ and $\sigma\sb{r}$) and the model parameters,
while keeping $\gamma$ fixed at 1.  We find that a non-informative
logarithmic prior on $\sigma\sb{r}$ handle the case with no correlated
noise better than a flat prior.  A log-flat prior is a scale-invariant
prior that has an equal probability per order of magnitude.  This is a
more convenient prior when the parameter may range over several orders
of magnitude \citep{Gregory2005BayesianBook}.  The only requirement is
that the parameter value must be positive.

\subsubsection{Results}
\label{sec:simresults}

To assess the quality of the inferences from the time averaging and
wavelet likelihood methods, we performed calibration tests, i.e.,
tests of the repeated-sampling (frequentist) performance of the
inferences, when applied across an ensemble of simulated datasets.

The first test computes a measure of the \emph{relative} accuracy of
the eclipse depth estimates (i.e., accuracy relative to the reported
uncertainty), also known as ``number-of-sigma'' statistic as described
by \citet{CarterWinn2009apjWavelets}.  The simulated datasets are
large enough that the marginal posterior PDFs for the eclipse depth
are typically nearly normal.  This motivates measuring relative
accuracy by the number of posterior standard deviations between the
best-fit value and the true value used for a simulation,
\begin{eqnarray}
{\cal N}\sb{p} = \frac{\hat p - p}{\sigma\sb{p}},
\label{eq:sigmastats}
\end{eqnarray}
where $p$ is the true eclipse depth used for the simulation, $\hat p$
is the best-fit estimate, and $\sigma_p$ is the standard deviation of
the marginal posterior for $p$.  If the marginal posterior were
normal, and if $\sigma_p$ were constant across the simulations, we
would expect ${\cal N}_p$ to have normal standard deviation with mean
$\langle{\cal N}_p\rangle$ zero, and standard deviation
$\sigma\sb{\cal N}$ unity.  In principle, a departure of $\langle{\cal
N}_p\rangle$ from zero would suggest a lack of accuracy of an
analysis, whereas a departure of $\sigma\sb{\cal N}$ from unity would
suggest an under- or overestimated precision ($\sigma\sb{\cal N}>1$ or
$\sigma\sb{\cal N}<1$, respectively).  In general, however, neither
condition rigorously holds; the posteriors are slightly non-normal,
and the value of $\sigma_p$ varies a bit.  As a result, we do not
expect ${\cal N}_p$ to have a standard normal distribution exactly
(even for many simulations).  Nevertheless, the mean and shape of the
${\cal N}_p$ distribution can reveal significant calibration failures
of inferences.

The second test examines the conditional coverage of marginal credible
regions for the eclipse depth.  For a set of simulations with
parameters $\theta$, we compute $C_Q(\theta)$, the coverage of
marginal credible regions for the eclipse depth that were computed to
contain a fraction $Q$ of the posterior probability.  That is, we fix
a size for the eclipse depth credible region to be tested (say,
$Q=0.683$ for a conventional ``1$\sigma$'' region), and we compute the
fraction of times the credible region contains the true eclipse depth
value.

Some caution is required in interpreting results of conditional
coverage tests.  Bayesian credible regions will not in general be
perfectly calibrated for fixed $\theta$.  Rather, Bayesian methods
produce regions with exact \emph{average} coverage, i.e., with
$\langle C_Q(\theta)\rangle = Q$ if one averages over the prior.
The bottom line of these considerations is that, for a set of
simulations with a fixed set of true parameter values, we do not
expect $C_Q(\theta)$ to equal $Q$ exactly.  But large departures from
$Q$ likely indicate problems with an inference procedure.  It is
possible, at least in principle, to more thoroughly verify calibration
of Bayesian MCMC algorithms \citep[e.g., to average over $\theta$, or
to consider all possible sizes of credible regions;
see][]{Cook2006jcgsBayesianValidation}, but we focus on simpler
conditional tests here.  For the coverage calculations reported below,
we used the 68.3\% marginal highest posterior density credible region.

\begin{figure}[tb]
\centering
\includegraphics[width=\linewidth, clip]{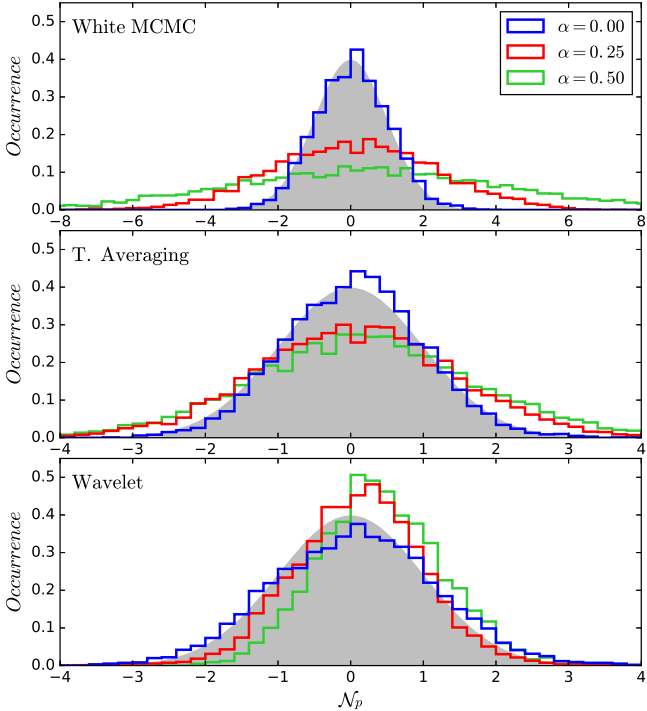}
\caption{Eclipse-depth histogram of the relative-accuracy statistic for
 synthetic-data simulations.  The top, middle, and bottom panels show
 the results for the white-MCMC, wavelet-likelihood, and
 time-averaging methods, respectively. The color code denotes the
 sample (see legend).  The background gray contour denotes a standard
 normal distribution for comparison.  All histograms are normalized
 such that the integral of each curve adds to one.}
\label{fig:syntheticResults}
\end{figure}

Figure \ref{fig:syntheticResults} shows the ${\cal N}\sb{\rm p}$
distributions for each method and dataset.  Table \ref{table:results}
shows the mean and standard deviation of ${\cal N}\sb{\rm p}$ and
coverage fraction $C\sb{Q=0.68}$ (i.e., the fraction of trials where
the 68\% HPD covers the true depth value).  The white analysis of the
uncorrelated-noise set serves as a control sample.  As expected, the
${\cal N}_p$ distribution for this case shows a negligible deviation
from zero, a standard deviation close to one, and a 68\%-HPD coverage
fraction of $\sim$68\%.  The correlated-noise runs reveal the failure
of the white analysis to account for correlated noise; as the
correlated-noise component increases, the mean and standard deviation
of ${\cal N}_p$ increase (suggesting a decrease of accuracy and
underestimated uncertainties), which is well correlated with the lower
coverage fraction.

The time-averaging method seems to improve the precision for the
correlated-noise runs with respect to those of the white analysis
(less underestimated uncertainties), as shown by the smaller
$\sigma\sb{\cal N}$.  Accordingly, the coverage fractions closer to
68\% indicate an improvement in the parameter estimation.  Note that
the time-averaging method does not affect the accuracy (with respect
to the white analysis).  Since all the data uncertainties are inflated
by a common scaling factor, the best-fitting solution does not change.
Therefore, the variation in $\langle{\cal N}_p\rangle$ is a
consequence of the variation in the precision and any possible
underlying correlation between precision and accuracy for the sample.

The wavelet method also seems to improve the parameter estimation of
the correlated samples with respect to the white analysis.  However,
this time the method seems to overestimate the uncertainties, as shown
by the coverage fractions greater than 68\% and the values of
$\sigma\sb{\cal N}$ smaller than one.  The values of $\langle{\cal
N}_p\rangle$ lie at the same level as those of the white analysis.

Overall, both the time-averaging and wavelet methods improve the
eclipse-depth estimation over a white analysis.  For a sample size of
3000 trials, the coverage uncertainty is roughly 2\% (from a root-$N$
estimate).  Since we carried out only a conditional study, there is an
additional error budget to consider.  Only a substantial mismatch
between coverage and credible region size would be evidence that there
is a problem.  The results of the wavelet analysis may be evidence of
a real coverage mismatch, but it is not at a level that would be
surprising for conditional {\vs} average coverage.  If real, this
emphasizes the challenges of retrieving reliable parameter estimates
from light curves affected by correlated noise, considering that we
generated the synthetic signal with wavelet function.

\begin{table}[ht]
\centering
\caption{Relative-accuracy Statistics and Coverage Fraction for Synthetic-data Simulations}
\label{table:results}
\begin{tabular*}{\linewidth}{@{\extracolsep{\fill}}lccc}
\tableline
\tableline
Estimation method & $\langle{\cal N}_p\rangle$ & $\sigma\sb{\cal N}$ &
                                          $C\sb{0.68}$ \\
\tableline
{\bf White MCMC}     \\
\hspace{0.3cm} ${\alpha=0.00}$  & 0.048   & 1.008   & 0.67 \\
\hspace{0.3cm} ${\alpha=0.25}$  & 0.091   & 2.230   & 0.33 \\
\hspace{0.3cm} ${\alpha=0.50}$  & 0.420   & 3.826   & 0.21 \\
{\bf Time Averaging} \\
\hspace{0.3cm} ${\alpha=0.00}$  & 0.044   & 0.954   & 0.71 \\
\hspace{0.3cm} ${\alpha=0.25}$  & 0.062   & 1.418   & 0.52 \\
\hspace{0.3cm} ${\alpha=0.50}$  & 0.177   & 1.581   & 0.48 \\
{\bf Wavelet}        \\
\hspace{0.3cm} ${\alpha=0.00}$  & 0.056   & 1.128   & 0.63 \\
\hspace{0.3cm} ${\alpha=0.25}$  & 0.111   & 0.897   & 0.74 \\
\hspace{0.3cm} ${\alpha=0.50}$  & 0.408   & 0.846   & 0.78 \\
\tableline
\end{tabular*}
\end{table}

\subsection{Synthetic-noise for Long-duration Simulation}
\label{sec:synthetic2}

Here we describe tests of the time-averaging method for datasets long
enough such that the eclipse duration lies at timescales where the
asymptotic approximation is still valid.  To do so, we replicate the
previous simulation (synthetic transit, white noise, and 1/$f$ noise
signals) for an observation lasting $\sim 20$ times the eclipse
duration (akin to a phase-curve observation).  We generate the light
curve with the same eclipse configuration and system flux as in
Section \ref{sec:synthetic2}, keeping the cadence (17,000 data points
total) and the value of $\sigma\sb{\omega}$ at 64.5 counts.  To
conserve the noise rms ratios at $\alpha=0.25$ and 0.5, we set
$\sigma\sb{r}=774$ and 1549 counts, respectively.

\begin{figure}[tb]
\centering
\includegraphics[width=\linewidth, clip]{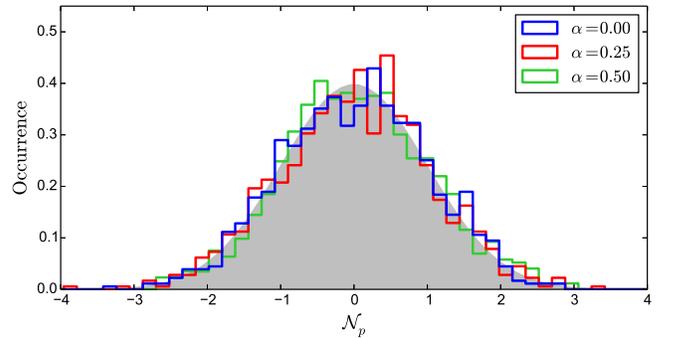}
\caption{Time-averaging eclipse-depth
 histogram of the relative-accuracy statistic for the long-duration
 (20 times the eclipse duration) synthetic-data simulations.  The
 color code denotes the sample (see legend).  The background gray
 contour denotes a standard normal distribution for comparison.  All
 histograms are normalized such that the integral of each curve adds
 to one.}
\label{fig:longrun}
\end{figure}

In this case, we find that the time-averaging $\beta$ scaling factors
accurately inflate the data uncertainties to account for the
time-correlated noise (Fig.\ \ref{fig:longrun}).

\subsection{Simulation with Spitzer-IRAC Noise}
\label{sec:semisynth3}

In this section we describe tests of the correlated-noise estimators
for real exoplanet signals from the Spitzer IRAC instrument, which is
more compelling than the previous test with synthetic data.  We select
two published phase-curve observations that are affected by correlated
noise, a 4.5 {\microns}
HD~209458\,b \citep{ZellemEtal2014apjHD209Phase} and a 3.6 {\microns}
WASP-14\,b \citep{WongEtal2015apjWASP14bSpitzerPhaseCurve} dataset.
The strength of the correlated noise in these two datasets is markedly
different.  The WASP-14\,b not only presents higher levels of
correlated noise (as reported by the time-averaging curves), but also
presents sporadic short-duration flux anomalies (dips) along the
observation.  Thus, these two datasets allow us to test the
correlated-noise estimators under true instrumental noise as detected
by the telescope, under two different correlated-noise regimes.  We
specifically selected phase-curve observations to remove the
astrophysical signals and trace the telescope systematics to the best
of our knowledge.

We processed the Spitzer BCD data to obtain raw light
curves using the Photometry for Orbits, Eclipses, and Transits (POET)
pipeline \citep{StevensonEtal2010natGJ436b,
 StevensonEtal2012apjHD149026b, StevensonEtal2012apjGJ436c,
 CampoEtal2011apjWASP12b, NymeyerEtal2011apjWASP18b,
 CubillosEtal2013apjWASP8b, CubillosEtal2014apjTrES1}.  
The POET pipeline involves bad-pixel masking (sigma rejection),
2-dimensional Gaussian fitting to determine the target location, and
interpolated aperture photometry to obtain raw light curves \citep[for
details see, e.g.,][]{CubillosEtal2014apjTrES1}.  We remove the first
couple hours of observation to avoid the time-dependent systematic.

We model the light curves using \citet{MandelAgol2002ApJtransits}
eclipse and transit models, a BLISS map model \citep[to account for
the intrapixel effect,][]{StevensonEtal2012apjHD149026b}, and and a
sinusoidal function \citep[for the phase-curve variation,
following][]{ZellemEtal2014apjHD209Phase}:
\begin{equation}
F(t) = 1 + c\sb{0} + c\sb{1} \cos(2\pi t) + c\sb{2} \sin(2\pi t),
\end{equation}
where $c\sb{0}$, $c\sb{1}$, and $c\sb{2}$ are the model fitting
parameters, and $t$ is the time of the observation (measured in
orbital phase).  To avoid degeneracy with the other fitting
parameters, we constrain $c\sb{0}$ by requiring $F(t\sb{0}) = 1$, with
$t\sb{0}$ the eclipse midpoint time.  We finally remove all
astrophysical variations from the signal by dividing out the
sinusoidal model and trimming the HD~209458\,b phase curves to the
span between the eclipse and transit (38.9 h long) and the WASP-14\,b
phase curve between the transit and eclipse (22.6 h long).  The
resulting light curves consist of flat curves containing only the
intra-pixel systematic variation and noise.  These curves are our
baseline to create the synthetic transit observations.

To construct the trial samples, we inject an eclipse curve
at random uniformly distributed times into the baseline, generating
3000 realizations for each dataset.  We adopt eclipse parameters
(duration, depth, ingress, and egress) similar to the observed values
for each dataset \citep{ZellemEtal2014apjHD209Phase,
WongEtal2015apjWASP14bSpitzerPhaseCurve}.

We analyze the data and outputs in the same manner as in
Section \ref{sec:simresults}.  Our fitting model includes an eclipse
and a BLISS-map model.  In practical terms, we found that many times
the MCMC for the wavelet method failed to converge or failed to fit
well the entire light curve.  This may be result of the wavelet noise
model attempting to overfit the transit curve, or because the wavelet
cannot appropriately model correlated-noise structure.  Thus, for this
analysis we trim the dataset to a window of 2.5 times the eclipse
duration, centered at the input midpoint.

\begin{figure}[tb]
\centering
\includegraphics[width=\linewidth, clip]{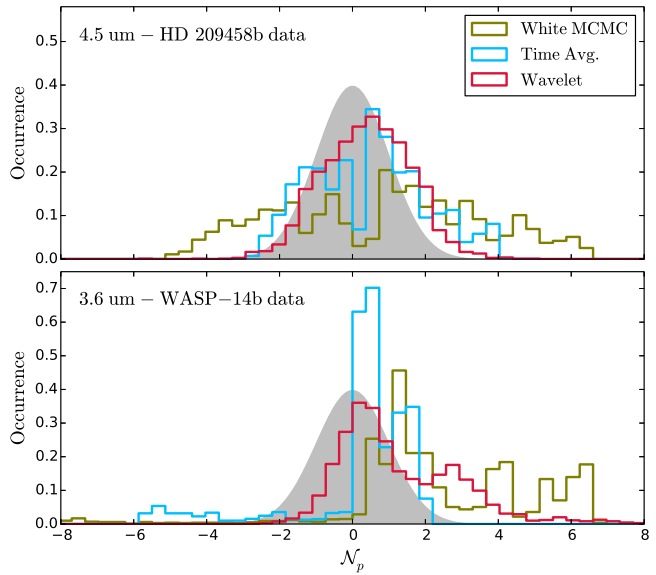}
\caption{Eclipse-depth histogram of the relative-accuracy statistic for
 real-noise simulations.  The top and bottom panels show the results
 for the HD~209458\,b and WASP-14\,b datasets, respectively. The color
 code denotes the parameter-estimation method (see legend).  The
 background gray contour denotes a standard normal distribution for
 comparison.  All histograms are normalized such that the integral of
 each curve adds to one.}
\label{fig:HD209simulation}
\end{figure}

\begin{table}[ht]
\centering
\caption{Relative-accuracy Statistics and Coverage Fraction for Real-noise Simulations}
\label{table:HD209results}
\begin{tabular*}{\linewidth}{@{\extracolsep{\fill}}lccc}
\tableline
\tableline
Estimation method    & $\langle{\cal N}_p\rangle$ & $\sigma\sb{\cal N}$ & $C\sb{0.68}$   \\
\tableline
{\bf 4.5 {\microns} HD~209458\,b} \\
\hspace{0.3cm} White MCMC           &  0.726      & 2.772   &  0.20 \\
\hspace{0.3cm} Time Averaging       &  0.444      & 1.553   &  0.46 \\
\hspace{0.3cm} Wavelet              &  0.428      & 1.176   &  0.54 \\
{\bf 3.6 {\microns} WASP-14\,b} \\                            
\hspace{0.3cm} White MCMC           &  0.720      & 5.901   &  0.15 \\
\hspace{0.3cm} Time Averaging       &  0.202      & 1.682   &  0.64 \\
\hspace{0.3cm} Wavelet              &  1.286      & 1.766   &  0.54 \\
\tableline
\end{tabular*}
\end{table}

Figure \ref{fig:HD209simulation} shows the resulting ${\cal N}\sb{\rm
p}$ histograms for each method and dataset.  The ${\cal N}\sb{\rm p}$
histograms for the WASP-14\,b dataset are noticeably more irregular
(non-Gaussian) than the histograms for the HD~2019458\,b dataset.
This may be result of the stronger correlated-noise systematics.  This
clearly complicates the interpretation of the ${\cal N}\sb{\rm p}$
statistics.  Thus, we rely mostly on the coverage-fraction statistic,
which is not affected by these nuances.

Table \ref{table:HD209results} presents the ${\cal N}\sb{\rm p}$
statistic and coverage-fraction results.  Again, in both cases, both
the time-averaging and wavelet methods improve the parameter
estimation with respect to the white analysis (coverage fractions
closer to 68\%).  However, none of the analyses completely correct the
eclipse depth estimation.  The low values of the coverage fractions
hint towards an underestimation of the uncertainties, a lack of
accuracy, or a combination of both.  The irregular shape of the ${\cal
N}\sb{\rm p}$ histograms, particularly for the WASP-14\,b dataset,
suggests that there are correlations between the accuracy and
precision for the trials, which would be expected given the stronger
correlated-noise component in the data.

\section{MULTI-CORE MARKOV-CHAIN MONTE CARLO (MC\sp{3}) CODE}
\label{sec:mc3}

We implemented and made available all of the discussed statistical
methods into the open-source Python package Multi-Core Markov-Chain
Monte Carlo (\href{https://github.com/pcubillos/MCcubed} {\mcc,
https://github.com/pcubillos/MCcubed}).  Unlike other exoplanet
model-fitting tools that are tailored to specific tasks, {\mcc} allows
the user to define the modeling function and, thus, it is a
general-purpose statistical package.  We developed the main bulk of
the code in Python, with several extensions written in C, combining
simplicity and high performance.  The code runs in multiple parallel
processors (through the built-in \texttt{multiprocessing} Python
package).  {\mcc} provides statistically-robust model optimization
(via Levenberg-Marquardt minimization) and credible-region estimation
(via MCMC sampling) routines.

The MCMC random sampling is done via the Metropolis Random Walk (MRW,
using multivariate Gaussian proposals), the Differential-Evolution
Markov-chain Monte Carlo \citep[DEMC,
][]{Braak2006DifferentialEvolution}, or the Snooker-updater DEMC
algorithms \citep{Braak2008SnookerDEMC}.  While the proposal step
sizes of the MRW are predetermined by the user and have to be manually
adjusted before each run, the DEMC algorithms automatically adjust the
scale and orientation of the proposal distribution.  To do so, DEMC
runs several chains in parallel, computing the proposed jump for a
given chain from the difference between the parameter states of two
other randomly selected chains.  As the chains converge toward the
posterior distribution, the proposal jumps will be mainly oriented
along the desired distribution and will have adequate scales.
Therefore, DEMC improves the MCMC efficiency in two ways: (1) it
increases the acceptance rate to optimal levels \citep[$\gtrsim
25$\%,][]{RobertsEtal1997} by better sampling the parameter space, and
(2) it eliminates the heuristic need for the user to adjust the
proposal jump scales.

The Metropolis-Hastings acceptance rule implements both
the ordinary likelihood function
(Eq.\ \ref{eq:samp-ind}) and the wavelet-based likelihood
(Eq.\ \ref{eq:wavelike}) using the fourth-order Daubechies
wavelet.  The priors can be bounded or unbounded uniform, log-scale
uniform, or Gaussian.
To assess that the MCMC is working properly, the code performs a
chain-converge test using the
\citet{GelmanRubin1992} statistics.  The code also produces several
plots to help visualize the results: trace, rms-vs.-bin-size,
marginal-posterior, and pairwise-posterior plots can indicate
non-convergence, multi-modal posteriors, parameter correlations,
correlated noise, or incorrect priors.  At the end of the MCMC run the
code returns the sampled posterior distribution of the parameters,
their best-fitting values, their 68\% HPD credible region, and the
acceptance rate of the MCMC.
The majority of the routines of this module derive from our POET
pipeline and, thus, have been thoroughly tested for years.

The core structure of {\mcc} consists of a central hub, which drives
the MCMC exploration, and the workers, which evaluate the model for
the given free parameters.  The hub and the worker processes are
connected through shared memory.  {\mcc} assigns one CPU to each
worker (i.e., one for each chain).  Each cycle (iteration) of the MCMC
comprises the following steps: (1) generate the proposal state (the
set of free parameters) for each chain, (2) evaluate the model for the
proposed state, and (3) compute the Metropolis ratio and accept/reject
the proposal state for each chain.

The {\mcc} code runs from both the shell prompt and the Python
interactive interpreter, and is available for Python 2.7 and Python 3.
The user can configure the MCMC run either through a configuration
file, command line arguments (prompt), and/or function arguments
(Python interpreter).  The minimum required inputs are the modeling
function, the data being fitted, and starting estimate values for the
free parameters.  As optional arguments, the user can supply the data
uncertainties, priors, and any extra arguments of the modeling
function (in a manner much like the
\texttt{scipy.optimize.leastq} routine).  Additionally, the package
allows the user to configure multiple features of the MCMC, e.g.:
number of chains, number of iterations, burn-in length, thinning
factor, etc.  The repository of the code includes a user manual and
guided examples.

\section{CONCLUSIONS}
\label{sec:conclusions}

Time-correlated noise is an important source of uncertainty for faint
signals such as exoplanet light curves.  Unless all systematics of the
data are well understood, the correlated noise must be taken into
account to obtain a reliable parameter estimation.  We have reviewed
three of the most widely used methods to assess time-correlated noise
in exoplanet time-series data: time averaging, residual permutation,
and wavelet-based likelihood, expanding the limited literature of
tests to assess the quantitative results of these techniques.  We focused
specifically on the case of Spitzer secondary-eclipse time-series
data.

We characterized the behavior of the time-averaging $\beta$ correction
factor at large bin sizes (the typical case for a transit
observation).  In this regime the assumed uncertainty
of the rms curve is no longer valid, since the posterior adopts
the form of a skewed inverse-gamma distribution.  We also found the
residual-permutation method unsound as a tool for quantifying
uncertainty in parameter estimates, because it does not produce
ensembles that mimic the behavior of independent draws from a
probability distribution.  The method is not supported by a
consistent statistical basis.  Finally, for the wavelet-likelihood
method we detected and corrected errors in the published equations
\citet{CarterWinn2009apjWavelets} and code (Appendix \ref{sec:errata}).

To quantitatively test the performance of these methods, we carried
out injection--retrieval simulations on synthetic eclipse light
curves, creating a large number of trials for each simulation.
We analyzed the results by (1) comparing the expectation and
standard deviation of the relative accuracy against a normal
distribution \citep[following ][]{CarterWinn2009apjWavelets}  and (2)
computing the 68\% coverage fraction (the fraction of trials where the
68\% credible region contained the injected parameter).
A correct Bayesian calculation would guarantee matching of the
  credible region probability and the average coverage (over the
  parameter space).  Note that our simulations used a single true
  parameter value.  A precise test of average coverage would require a
  substantial amount of computing time.  Thus, our conditional coverage
  tests would indicate a problem with a method only if the coverage
  differed substantially from the credible level.

In our first simulation, we tested the case when the time-correlated
noise has a power spectral density of the form $1/f$.  Both the
time-averaging and the wavelet-likelihood methods improved the
eclipse-depth estimations over a white MCMC analysis.  In this
simulation the wavelet analysis is expected to perform well, since the
wavelet precisely assumes a noise component with a $1/f$ power
spectral density.  We found small differences between the estimated
conditional coverage and the credible levels.  These diffrences are
consistent with expectations, given both the limited precision from
the size of the simulations, and the conditional nature of the tests.
We also note that the performance of the time-averaging correction can
be further improved if the total observation time is much longer than
the eclipse duration (as in a phase-curve observation).  This arises
from the lower signal to noise of the ${\rm rms}\sb{N}$ curve at large
bin sizes.

In a further simulation we generated eclipse light-curve samples by
injecting an eclipse signal into real Spitzer 3.6~{\micron} and
4.5~{\micron} IRAC time-series datasets, two sets with notoriously
different correlated-noise signals.  This experiment allowed us to
assess the performance of the time-correlated estimators without
assuming a specific shape of the time-correlated signal.  Both the
time-averaging and the wavelet-likelihood methods significantly
improved the uncertainty estimations compared to a white MCMC
analysis, raising the coverage fraction from 15\%--20\% to
$\sim$45--65\%.  However, they are not perfect, as the coverage
fractions are still lower than the expected 68\%, suggesting a lack of
accuracy and (or) underestimated uncertainties.

In conclusion, it is always better to try to determine the best
possible model for the systematics than simply inflating the parameter
uncertainties (as in time averaging).  However, these sub-optimal
noise estimators are better than ignoring time-correlated noise.
Luckily, the continuous development of advanced data analysis
techniques like Gaussian Processes, Independent Component Analysis, or
kernel regression decorrelation \citep[see
e.g.,][]{IngallsEtal2016ajRepeatability} will help to improve the best
practices required to extract exoplanet data.

Lastly, we presented the open-source Python package Multi-Core
Markov-Chain Monte Carlo, available
at \href{https://github.com/pcubillos/MCcubed}
{https://github.com/pcubillos/MCcubed}.  {\mcc} implements all of the
statistical routines described in this paper, allowing the user to
estimate best-fitting model parameters and their credible region,
while letting the user provide the modeling function.  By releasing
our code to the community, we hope not only to provide access to the
routines discussed here, but also to encourage researchers to consider
open development and cross-validation of the software tools used in
the field.

\acknowledgments

We thank Rebekah Dawson for useful conversations.
We thank the referee for comments that
significantly improved the quality of the paper.  We thank
contributors to AstroPy \citep{Astropy2013aaAstroPy}, SciPy,
Matplotlib \citep{Hunter2007ieeeMatplotlib}, the Python Programming
Language, and the free and open-source community.  PC was partly
supported by the Fulbright Program for Foreign Students.  JB was
partly supported by the NASA Earth and Space Science Fellowship
Program, grant NNX12AL83H.  Part of this work is based on observations
made with the Spitzer Space Telescope, which is operated by the Jet
Propulsion Laboratory, California Institute of Technology under a
contract with NASA.  Support for this work was provided by NASA
through an award issued by JPL/Caltech and through the NASA Science
Mission Directorate's Astrophysics Data Analysis Program, grant
NNX13AF38G, and its Planetary Atmospheres Program, grant NNX12AI69G.

\facility{Spitzer(IRAC)}

\software{{\mcc} (\href{https://github.com/pcubillos/MCcubed}
{https://github.com/pcubillos/MCcubed}), Python}


\begin{appendices}

\section{BAYESIAN CREDIBLE REGION}
\label{sec:CredRegion}

In the Bayesian context, given the posterior probability density,
$p(\theta|\yvec)$, of a parameter, $\theta$, given the dataset,
$\yvec$, the highest posterior density region (or credible
region), $R$, is defined by
\begin{equation}
C = \int\sb{R} {\rm d}\theta\;p(\theta|\yvec)
\end{equation}
where C is the probability contained in the credible region.  The
region $R$ is selected such that the posterior probability of any
point inside $R$ is larger than that of any point outside.

In practice, to calculate the credible region, one constructs a
histogram of the sampled posterior distribution (normalized such that
the sum equals one) and sorts the bins in descending order.  Then one
sequentially adds the values of $p$ until reaching $C$.  The
credible-region boundaries are given by the smallest and largest
values of $\theta$ for the samples considered in the sum, if the
region is contiguous.

\section{WLS {\vs}\ GLS EXAMPLES}
\label{sec:OLSGLSexamples}

To gain insight into the difference between {\it weighted least
squares} estimates (WLS, those considering Eq.\ (\ref{eq:samp-ind}))
and {\it generalized least squares} estimates (GLS, those considering
Eq.\ (\ref{eq:corrPDF})), we consider examples with a simple
correlated noise model: AR(1) autoregressive noise, for regularly
sampled data.  Our treatment adapts analyses
by \citet{Zellner1971BayesianInference}
and \citet{SiviaSkilling2006Bayesian} on related models.  In this
model, the conditional expectation (regression) of the noise for
sample $i$ is proportional to the previous noise value; the actual
value of the noise is the sum of this expectation and a new
\emph{innovation} contribution, $\nu_i$:
\begin{equation}
\eps_i = \phi \eps_{i-1} + \nu_i,
\label{eps-AR1}
\end{equation}
where $\phi$ is the autoregression parameter.  The innovations are
independent, with zero-mean normal PDFs with standard deviation $s$.
The overall model for $f(\params)$ is a \emph{hidden Markov model}
(HMM): ``Markov'' indicating that the prediction for the noise at time
$t_i$ depends only on the noise at the previous time, and not on the
whole noise history; and ``hidden'' because $\eps_i$ is not directly
observed (as it would be in a standard AR(1) model), rather, $y_i$ is
observed, mixing uncertain model and noise contributions.

The AR(1) model enables recursive construction of the joint
distribution for the noise.  The model specifies independent normal
PDFs for the $\nu_i$ terms, so the goal is to express the $\eps_i$
values entirely in terms of $\nu_i$ values.  The probability for the
first noise sample, $\eps_1$, is slightly complicated by the fact that
it depends on innovations at times before there is data.  However,
$\eps_i$ is a linear sum of terms that are each zero-mean normal, so
it must itself have a normal PDF, with variance given by the sum of
the variances of its contributions.  Writing $\eps_{i-1} = \nu_{i-1}
+ \phi_{i-1}$, and recursing, we find
\begin{equation}
\eps_1 = \sum_{j=0}^{\infty} \phi^j\nu_{1-j}.
\label{eps1-AR1}
\end{equation}
The standard deviation of each term is $\phi^j s$, so the sum of the
variances is
\begin{align}
\sigma_{\eps}^2 = s^2\sum_{j=0}^\infty \phi^{2j} = \frac{s^2}{1 - \phi^2},
\label{sig-eps}
\end{align}
provided that $|\phi| < 1$.  The
marginal PDF for $\eps_i$ at any time is a zero-mean normal with this
variance; the noise time series is thus \emph{stationary} (with the
same marginal distribution at each time).

We can write the joint PDF for all noise values in terms of factors
that condition on the previous history:
\begin{equation}
  p(\evec) = p(\eps_1)\, p(\eps_2|\eps_1)\, p(\eps_3|\eps_{1:2}) \cdots p(\eps_n|\eps_{1:n-1}),
\label{eps-fac}
\end{equation}
with $\eps_{i:j} = (\eps_i,\ldots,\eps_j)$.  Given the Markov property
of the AR(1) model, the joint PDF simplifies to
\begin{equation}
p(\evec) = p(\eps_1) \prod_{i=2}^n p(\eps_i|\eps_{i-1}).
\label{evec-AR1}
\end{equation}
Equation~\ceq{eps-AR1} implies that $p(\eps_i|\eps_{i-1})$ is the
probability that $\nu_i = \eps_i - \phi\eps_{i-1}$.  The factors
appearing in Eq.\ (\ref{evec-AR1}) are thus
\begin{equation}
p(\eps_1) = \frac{1-\phi^2}{s\sqrt{2\pi}} e^{-\eps_1^2/2s^2}
\label{eps1-pdf}
\end{equation}
and
\begin{equation}
p(\eps_i|\eps_{i-1}) = \frac{1}{s\sqrt{2\pi}}
  \exp\left[-\frac{1}{2s^2}(\eps_i - \phi\eps_{i-1})^2\right].
\label{deps-pdf}
\end{equation}

The observation equation, Eq.\ (\ref{y-fe}), indicates that the probability
for the data, $\yvec$, is the probability that the noise values take
on the values $\eps_i = y_i - f_i(\params)$.  Let $r_i(\params) \equiv
y_i - f_i(\params)$ denote the residuals from adopting the model with
parameters $\params$.  Then the PDF for the data can be written
\begin{equation}
p(\yvec|\params) = \frac{(1-\phi^2)^{1/2}}{s^n (2\pi)^{n/2}} e^{-Q(\theta)/2s^2},
\label{AR1-pdf-Q}
\end{equation}
with
\begin{align}
Q(\params)
  &= (1-\phi^2) r_1^2 + \sum_{i=2}^n (r_i - \phi r_{i-1})^2 \nonumber\\
  &= \sum_{i=1}^n r_i^2 + \phi^2 \sum_{i=2}^{n-1} r_i^2
     - 2\phi \sum_{i=2}^n r_i r_{i-1}.
\label{Q-resid}
\end{align}
The first term---the sum of squared residuals---is just the
``$\chi^2$'' term that appears in WLS (see Eq.\ (\ref{eq:samp-ind})).  When
$\phi \ne 0$, AR(1) noise correlations introduce new contributions to
$Q(\params)$, including a term resembling a lag-1 autocorrelation.
  These terms correspond to changes in the model basis
projections entailed by the correlations in a GLS analysis, versus a
WLS analysis.

\subsection{Example: Constant Signal}

The simple case of a constant signal model of unknown amplitude,
$f(t;\mu) = \mu$, is analytically tractable and is illuminating.
Substituting $r_i = y_i - \mu$ and minimizing $Q(\mu)$ leads to the
GLS estimate
\begin{equation}
\hat{\mu} = \frac{w_n \bar{y} + w_2 (y_1+y_n)/2}{w_n + w_2},
\label{mu-hat}
\end{equation}
where $\bar{y}$ is the sample mean, $\bar{y} \equiv (1/n)\sum_i y_i$,
and we have defined weights $w_n = n(1-\phi)$ and $w_2 = 2\phi$.  When
$\phi=0$ (independent noise), $\hat{\mu}$ is just the sample mean.
Otherwise, $\hat{\mu}$ is a weighted average of the full sample mean,
and the average of the first and last (i.e., the most widely
separated) samples.  As $\phi$ approaches unity (strongly positively
correlated noise), GLS instructs us to just average the most widely
separated samples.  In contrast, the WLS estimate is always the full
sample mean.  The WLS and GLS estimates thus will differ, not just in
the uncertainties they assign to the mean, but also in the actual
values of the estimates.

$Q(\mu)$ is quadratic in $\mu$, so the likelihood function is a
Gaussian function in $\mu$.  The reciprocal of the second derivative
of $Q(\mu)/s^2$ at $\hat{\mu}$ gives the squared standard deviation of
this Gaussian,
\begin{equation}
\sigma_\mu^2 = \frac{s^2}{n(1-\phi)^2 - 2\phi(1-\phi)}.
\label{sig-mu}
\end{equation}
When $\phi=0$, we have $\sigma_\mu = s/\sqrt{n}$, the familiar
``root-$n$'' result.  As $\phi$ approaches unity, the denominator
decreases toward zero, and the uncertainty in $\mu$ grows.
Roughly
speaking, growing positive correlation decreases the effective sample
size, inflating uncertainties.  This motivates approaches like time
averaging that attempt to account for correlation merely by inflating
uncertainties.  But such approaches do not account for the effect of
correlations on the actual value of a finite-sample estimate.

\subsection{Example: Constant Baseline with One Dip}

The effect of correlations on parameter estimates depends on the
extent to which the correlations may mimic or distort the projections
of the data onto the model components.  When a model has components
that vary slowly with respect to the correlation scale, the main
effect of correlations is to change the effective sample size.  But
when a model has temporally localized components, correlations can
significantly affect, not just the uncertainty scale, but also the
best-fit parameter values.

\begin{figure}
\centering
\hspace*{0.77cm}\includegraphics[width=0.89\linewidth]{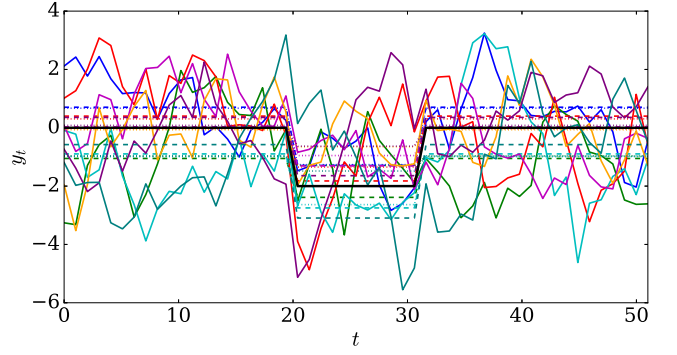}
\caption{Eight sampled time series from the baseline-dip model with AR(1)
  noise. Curves of matching color connect the simulated data (solid),
  the GLS best-fit points (dashed), and the WLS best-fit points
  (dotted). The solid black curve shows the true (noiseless)
  function.}
\label{fig:series}
\end{figure}

To illustrate this, we used simulated AR(1) noise and the GLS
likelihood function to model data generated from a baseline signal of
amplitude $a$, with a localized dip of depth $\delta$.  We took the
dip location and width to be known.  For the illustration we report
here, we simulated 51 observations with true parameter values $\params
= (a,\delta) = (0, 2)$, with the dip spanning 10 samples in the middle
of the time series.  The noise was generated with an innovation
standard deviation $s=1$, and $\phi=0.8$, producing data with
autocorrelation time scales $\sim 5$.  Figure~\ref{fig:series}
displays examples of the simulated data and WLS and GLS best-fit
function estimates.  It is visually apparent that the WLS and GLS
estimates sometimes differ.

\begin{figure}
\centering
\includegraphics[width=\columnwidth]{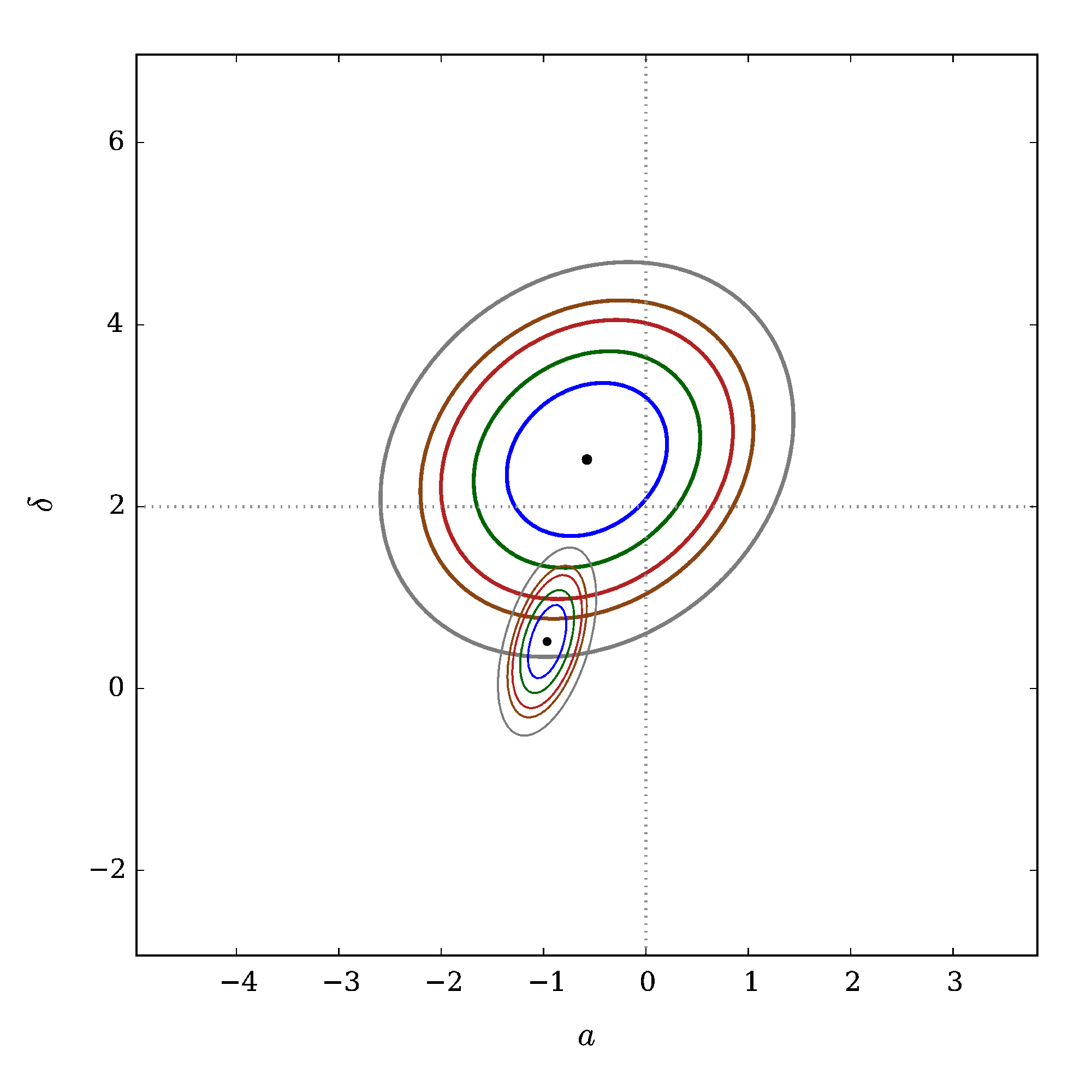}
\caption{
  Contours of the posterior PDF for $(a,\delta)$, from GLS (larger,
  thicker contours) and WLS (smaller, thinner
  contours) analyses.  From inside to outside, the contours bound
  highest posterior density regions with 50\%, 75\%, 90\%, 95\%, and
  99\% of the posterior probability. Dots indicate the
  modes. Crosshairs indicate the true parameter
  values.}  \label{fig:contours}
\end{figure}

Figure~\ref{fig:contours} shows contours of the posterior PDFs for
$(a,\delta)$ from a representative simulation where the WLS and GLS
estimates differ; here the WLS best-fit estimate is just outside of
the 98\% GLS credible region.  Even when the WLS and GLS best-fit
estimates did not differ too dramatically in, the WLS likelihood
function not only has an incorrect uncertainty scale (which one might
hope to fix via inflation), but does not correctly capture the shape
of the PDF (i.e., the correlation between $a$ and $\delta$ estimates).

The main message of these examples is that noise correlation not only
can inflate uncertainties; it can also corrupt parameter estimates,
particularly when parameters of interest pertain to temporally
localized structure in the model, for which noise correlations can
significantly change the data projections needed for accurate
inference.  Methods that seek to account for correlations only by
inflating parameter uncertainties are at best suboptimal (producing
larger estimation errors than could be achieved with a good correlated
noise model), and can sometimes be significantly misleading.

\section{STANDARD-DEVIATION UNCERTAINTY}
\label{sec:StdUncert}

The uncertainty of a parameter estimate in a problem with a fixed
model dimension (number of parameters) and growing sample size
typically decreases asymptotically at the {$\sqrt{M}$} rate.  That is,
for estimating a Gaussian mean from {$M$} samples with the standard
deviation $\sigma$, which is known, the uncertainty is
{$\sigma/\sqrt{M}$}.  However, this result says nothing about the actual
{\em size} of the uncertainty at any particular sample size.  When
$\sigma$ is unknown, it becomes the target of estimation, instead of
(or in addition to) the mean.  Here, we elaborate on the derivation
of the uncertainty for the standard deviation of a Gaussian.  The
derivation uses the Laplace approximation for a normal standard
deviation and its uncertainty, i.e., it finds a Gaussian distribution
with a peak and curvature matching the marginal probability density
function.

Given a normal distribution of values with unknown mean $\mu$ and
standard deviation $\sigma$, let $b\sb{i}$ be the means for a sample
of $M$ groups of samples (``bins'') drawn from this distribution.
The sample mean, $\bar{b}$, and the sample variance, $s\sp{2}$,
are defined as usual:
\begin{equation}
\label{Aeq:meanstd}
\bar{b} = \frac{1}{M}\sum\sb{i} b\sb{i}, \qquad
s\sp{2} = \frac{1}{M}\sum\sb{i} (b\sb{i} - \bar{b})\sp{2}.
\end{equation}

If the residual $r\sb{i} = b\sb{i} - \bar{b}$ and $r\sp{2} = \sum\sb{i} r\sb{i}\sp{2}$, then the sample
variance becomes $s\sp{2} = r\sp{2}/M$.

The likelihood function for our normal distribution with ($\mu$,
$\sigma$) is:
\begin{eqnarray}
\label{Aeq:likelihood}
\cal{L}(\mu, \sigma) & = & \prod\sb{i} p\,(b\sb{i}|\mu, \sigma) \nonumber \\
                    & = & \prod\sb{i} \frac{1}{\sigma\sqrt{2\pi}}
                 \exp\left[-\frac{(b\sb{i}-\mu)\sp{2}}{2\sigma\sp{2}}\right],
\end{eqnarray}

\noindent so the likelihood can be written is terms of $\bar{b}$ and $r$ as:
\begin{eqnarray}
\label{Aeq:likelihood2}
\cal{L}(\mu,\sigma)
           & = & \frac{1}{\sigma\sp{M} (2\pi)\sp{M/2}}
                 \exp\left(-\frac{r\sp{2}}{2\sigma\sp{2}}             \right) 
                 \exp\left(-\frac{M(\mu-\bar{b})\sp{2}}{2\sigma\sp{2}}\right).
\end{eqnarray}

To estimate $\mu$ and $\sigma$, we will adopt a flat prior for $\mu$
and a log-flat prior for $\sigma$, corresponding to $p(\sigma) \propto
1/\sigma$. Then, the joint posterior probability $p(\mu,\sigma|D)$ for
$\mu$ and $\sigma$, given the data $D$ is:

\begin{equation}
\label{eq:posterior1}
p(\mu,\sigma|D) \propto p(\sigma) \times \cal{L}(\mu,\sigma),
\end{equation}
with $p(\sigma)$ the prior probability on $\sigma$.

\begin{eqnarray}
\label{eq:posterior2}
p(\mu,\sigma|D) 
 & \propto & \frac{1}{\sigma\sp{M+1}}  
             \exp\left(-\frac{r\sp{2}}{2\sigma\sp{2}}\right)
             \exp\left(-\frac{M(\mu-\bar{b})\sp{2}}{2\sigma\sp{2}}\right).
\end{eqnarray}

Calculate the marginal posterior density for $\sigma$ by
integrating over $\mu$:
\begin{eqnarray}
\label{eq:margsigma}
p(\sigma|D)
      \propto  \int  \frac{{\rm d}\mu}{\sigma\sp{M+1}}
                 \exp\left( -\frac{r\sp{2}}{2\sigma\sp{2}} \right) 
                 \exp\left( -\frac{M(\mu-\bar{b})\sp{2}}{2\sigma\sp{2}}\right)
\end{eqnarray}

The $\mu$ dependence is in the last exponential factor, a Gaussian that
integrates to $\sigma\sqrt{2\pi}$.  We denote the result as $f(\sigma)$:
\begin{eqnarray}
\label{eq:margsigma2}
\label{eq:fSigma}
p(\sigma|D) \propto \frac{1}{\sigma\sp{M}}
                   \exp\left( -\frac{r\sp{2}}{2\sigma\sp{2}} \right)
                   = f(\sigma).
\end{eqnarray}

We estimate $\sigma$ with its mode, $\hat\sigma$, which maximizes
$f(\sigma)$. The first derivative of $f(\sigma)$ is:
\begin{equation}
f'(\sigma) = f(\sigma)\left(\frac{r\sp{2}}{\sigma\sp{3}} - 
                            \frac{M}{\sigma} \right),
\label{eq:Tom15}
\end{equation}
so that setting $f'(\hat\sigma) = 0$ gives $\hat\sigma = r/\sqrt{M} = s$,
as one might expect.

For a simple estimate of the uncertainty, let's consider a Gaussian
approximation with mean at $\hat\sigma$.  The curvature (second
derivative) of $\sigma$ at $\hat\sigma$:
\begin{equation}
\label{eq:Tom16}
f''(\sigma) =
f'(\sigma) \left(\frac{r\sp{2}}{\sigma\sp{3}} - \frac{M}{\sigma}\right) +
 f(\sigma) \left(\frac{M}{\sigma\sp{2}} - \frac{3r\sp{2}}{\sigma\sp{4}}\right),
\end{equation}
determines the standard deviation.  When $f(x)$ is of the form of a
normal distribution with mean $m$ and standard deviation $w$, it is
easy to show that $f''(m) = -f(m)/w\sp{2}$.
So, if $\delta$ is the standard deviation for $\sigma$, in the normal
approximation, that matches the curvature at the peak, we have
$\delta\sp{2} \approx - f(\hat\sigma)/f''(\hat\sigma)$.  Evaluating
Equation (\ref{eq:Tom16}) at $\hat\sigma$, the first term vanishes
(since $f'(\hat\sigma)=0$), and the remaining term gives an
approximate standard deviation of:
\begin{eqnarray}
\label{eq:Tom17}
\delta \approx \frac{r}{M\sqrt{2}} = \frac{s}{\sqrt{2M}}.
\label{delta-approx}
\end{eqnarray}

So, the mean and standard deviation sum for $\sigma$ for large $M$ is:
\begin{eqnarray}
\label{eq:Tom18}
\sigma = s \pm \frac{s}{\sqrt{2M}}.
\label{sigma-est}
\end{eqnarray}

\section{Wavelet-likelihood Errata}
\label{sec:errata}

This section reports three erratas found both in the published
article of \citet{CarterWinn2009apjWavelets} and its associated ASCL
code.

First, in the Likelihood equation, Eqs.\ (32) and (41)
of \citet{CarterWinn2009apjWavelets}, the index for the scale, $m$,
should start from 1 instead of 2.  In this case the ASCL code has the
correct value.

Next, the variance of the scaling coefficient in the ASCL code for
$\gamma=1$, equation (34) of the paper, is missing the factor
$2\sp{-\gamma}=2\sp{-1}$.  The corrected equation should read:
\begin{equation}
\sigma\sb{S}\sp{2} = \frac{\sigma\sb{r}\sp{2}}{4\ln{2}} +
                    \sigma\sb{\omega}\sp{2}.
\end{equation}

The expression for the scaling coefficient for $\gamma\ne1$ also seems
to be wrong in the ASCL code.  We can compute the variance of the
wavelet coefficients following equation (37)
of \citet{Wornell1993ieeeWavelet}:
\begin{equation}
\langle \epsilon^{m}_{n}\epsilon^{m}_{n} \rangle = \frac{2^{-m}}{2\pi}\int^{\infty}_{-\infty}
      \frac{\sigma^{2}_{x}}{|\omega|^\gamma}|\Psi(2^{-m}\omega)|^{2}{\rm d}\omega.
\end{equation}
Assuming an ideal bandpass ---i.e., eq.\ (3) of
\citet{Wornell1993ieeeWavelet}--- and with a change of variable, $u =
2^{-m}\omega$, we reproduce  eq.\ (24) of
\citet{CarterWinn2009apjWavelets}:
\begin{eqnarray}
\langle \epsilon^{m}_{n}\epsilon^{m}_{n} \rangle & = &
            2^{-\gamma m} \frac{1}{2\pi}\int^{\infty}_{-\infty}
              \frac{\sigma^{2}_{x}}{|u|^\gamma}|\Psi(u)|^{2}{\rm d}u. \\
                    & = & 2^{-\gamma m} \frac{2}{2\pi}\int^{2\pi}_{\pi}
                          \frac{\sigma^{2}_{x}}{u^\gamma}{\rm d}u.  \\
                   & = & 2^{-\gamma m} \frac{\sigma^{2}_{x}}{\pi^\gamma}
                         \frac{[2^{1-\gamma}-1]}{1-\gamma}
                     \equiv 2^{-\gamma m} \sigma_r^2, 
\end{eqnarray}
Analogously, the variance for the scaling coefficient assuming
 an ideal bandpass ---i.e., eq.\ (10) of
\citet{Wornell1993ieeeWavelet}:
\begin{eqnarray}
\langle {\bar\epsilon}^{m}_{n}{\bar\epsilon}^{m}_{n} \rangle & = &
         2^{-\gamma m} \frac{1}{2\pi}\int^{\infty}_{-\infty}
         \frac{\sigma^{2}_{x}}{|\omega|^\gamma}|\Phi(\omega)|^{2}{\rm d}\omega\\
   & = & 2^{-\gamma m} \frac{2}{2\pi}\int^{\pi}_{0}
         \frac{\sigma^{2}_{x}}{\omega^\gamma}{\rm d}\omega \\
   & = & 2^{-\gamma m} \frac{\sigma^{2}_{x}}{\pi^\gamma} \frac{1}{1-\gamma}
   \equiv 2^{-\gamma m} \sigma^{2}_{r} \frac{1}{2^{1-\gamma}-1}. \label{eq:vars2}
\end{eqnarray}

This indicates that $g(\gamma) = 1/(2^{1-\gamma}-1)$ for $\gamma\ne1$, the
inverse of the value given in the ASCL code from
\citet{CarterWinn2009apjWavelets}.  The same result can be derived
from equations (16) and (17) of \citet{FadiliBullmore2002}.  These
derivations of $\langle \epsilon^{m}_{n}\epsilon^{m}_{n} \rangle$ and
$\langle {\bar\epsilon}^{m}_{n}{\bar\epsilon}^{m}_{n} \rangle$ are not
valid for $\gamma=1$; in fact, Eq. (\ref{eq:vars2}) diverges to
$+\infty$ from the left and to $-\infty$ from the right as we approach
$\gamma=1$. Then, how 
can one get to $g(\gamma=1)=1/2\ln2$?.

Lastly, Section 4.1 of \citet{CarterWinn2009apjWavelets} mentions that
they used a dataset of 1024 elements, and that their DWT produced 1023
wavelet coefficients and 1 scaling coefficient (implying $N\sb{0}=1$).
This is inconsistent with the wavelet used (a fourth-order Daubechies
wavelet), for which $N\sb{0}=2$.  This wavelet's DWT returns 2 scaling
coefficients and 1022 wavelet coefficients (for the given dataset).
The ASCL code is also suited to perform a likelihood calculation
assuming $N\sb{0}=1$, resulting in each likelihood term having an $m$
value offset by 1.

\end{appendices}
\end{document}